\documentclass[10pt,aps,twocolumn,groupedaddress,superscriptaddress,amsmath,amssymb,prb,longbibliography]{revtex4-2}
\usepackage{mathtools,amsmath,amsxtra}
\usepackage{amssymb}
\usepackage{physics}
\usepackage[english]{babel}
\usepackage{dsfont}
\usepackage{graphicx}
\usepackage{dcolumn}
\usepackage{bm}
\usepackage{bbold}

\usepackage{bigints}
\usepackage[usenames,dvipsnames]{xcolor}
\usepackage{tikz}
\usetikzlibrary{decorations.pathmorphing}
\usetikzlibrary{arrows.meta}
\usepackage{bm}
\usepackage{braket}
\usepackage[normalem]{ulem}
\usepackage{lipsum}
\usepackage{subcaption}
\usepackage{mathrsfs}
\usepackage[colorlinks,citecolor=Blue,linkcolor=Red, urlcolor=Blue]{hyperref}
\begin{document}

\title{Mitigating decoherence in molecular spin qudits}

\author{Leonardo Ratini}
\thanks{These authors contributed equally to this work.}
\author{Giacomo Sansone}
\thanks{These authors contributed equally to this work.}
\affiliation{Universit\`a di Parma, Dipartimento di Scienze Matematiche, Fisiche e Informatiche, I-43124 Parma, Italy} 
\affiliation{INFN–Sezione di Milano-Bicocca, gruppo collegato di Parma, 43124 Parma, Italy}

\author{Elena Garlatti}
\affiliation{Universit\`a di Parma, Dipartimento di Scienze Matematiche, Fisiche e Informatiche, I-43124 Parma, Italy} 
\affiliation{INFN–Sezione di Milano-Bicocca, gruppo collegato di Parma, 43124 Parma, Italy}
\affiliation{Consorzio Interuniversitario Nazionale per la Scienza e Tecnologia dei Materiali (INSTM), I-50121 Firenze, Italy}
\author{Francesco Petiziol}
\affiliation{Institut für Theoretische Physik, Technische Universität Berlin, Hardenbergstrasse 36, 10623 Berlin, Germany}
\author{Stefano Carretta}
\email{stefano.carretta@unipr.it}
\author{Paolo Santini}
\email{paolo.santini@unipr.it}
\affiliation{Universit\`a di Parma, Dipartimento di Scienze Matematiche, Fisiche e Informatiche, I-43124 Parma, Italy} 
\affiliation{INFN–Sezione di Milano-Bicocca, gruppo collegato di Parma, 43124 Parma, Italy}
\affiliation{Consorzio Interuniversitario Nazionale per la Scienza e Tecnologia dei Materiali (INSTM), I-50121 Firenze, Italy}

\begin{abstract}

Molecular nanomagnets are quantum spin systems potentially serving as qudits for future quantum technologies thanks to their many accessible low-energy states. At low temperatures, the primary source of error in these systems is pure dephasing, caused by their interactions with the bath of surrounding nuclear spins degrees of freedom. Most importantly, as the system’s dimensionality grows going from qubits to qudits, the control and mitigation of decoherence becomes more challenging. Here we analyze the characteristics of pure dephasing in molecular qudits under spin-echo sequences. We use a realistic description of their interaction with the bath, whose non-Markovian dynamics is accurately computed by the cluster correlation expansion technique. First, we demonstrate a necessary and sufficient condition to prevent the decay of coherence with time, also introducing a parameter to quantify the deviation from such ideal condition.  We illustrate this with two paradigmatic systems: a single giant spin and a composite antiferromagnetic spin system. We then advance a proposal for optimized nanomagnets, identifying key ingredients for engineering robust qudits for quantum technologies.
\end{abstract}

\maketitle

\twocolumngrid 

\section{Introduction}
\label{sec:intro}

Quantum computation is a milestone with the potential to revolutionize various fields, from materials design to the optimization of complex processes. The hardest challenge in achieving practical quantum advantage lies in mitigating noise-induced errors. Even the most mature platforms, such as those based on superconducting transmon qubits or trapped ions, belong to the category of noisy intermediate scale quantum (NISQ) devices. 
These systems are constrained by limited relaxation ($T_1$) and dephasing ($T_2$) times, which restrict the number of quantum gates that can be reliably executed due to error accumulation. Consequently, minimizing the depth of quantum circuits is critical for implementing algorithms effectively on NISQ devices.
For example, in quantum computation, different algorithms \cite{Abrams1999,Du2010,Smart2021}, such as the Variational Quantum Eingensolver (VQE) \cite{Peruzzo2014,McClean2016}, have been proposed to address electronic structure problems. 
Despite significant efforts to minimize the resource requirements for implementing this algorithm \cite{Barkoutsos2018,Romero2018,Grimsley2019,Materia2024}, simulating challenging prototypical systems remains out of reach.
To overcome noise-related limitations and achieve fault-tolerant quantum computation, the implementation of quantum error correction (QEC) is necessary.
The core principle of QEC is expanding the Hilbert space beyond the conventional two-dimensional space of a single \textit{qubit}. This is achieved by replacing two-level qubits with more complex logical units composed of multiple physical qubits \cite{Devitt2013}.
However, this approach is significantly limited by the large resource demands of multi-qubit codes, particularly due to the necessity of non-local gates operating on distinct physical units, which makes QEC highly challenging \cite{Terhal2015}. This difficulty motivates the exploration of alternative paradigms for implementing QEC.
One such approach involves expanding the Hilbert space by encoding the logical qubit into a multi-level physical system (\textit{qudit}) \cite{Gottesman2001,Pirando2008,Cafaro2012,Albert2020}.
Whether applied to QEC, quantum computation, or quantum simulation, one of the primary advantages of qudits over traditional qubits lies in their ability to reduce the number of distinct physical units and two-body gates required for operations. This is because certain quantum operations that necessitate multiple gates in qubit-based systems can often be executed with just a single gate in qudit-based systems \cite{Roca-Jerat_2025,Chicco_2024}. As a result, qudits not only lower the overall gate count but also enhance computational efficiency, making them a promising alternative for scalable quantum information processing.

In this regard, molecular spin systems and more specifically molecular nanomagnets (MNMs), represent an attractive platform due to the presence of many low-energy spin states, both nuclear and electronic ones, which can be exploited to encode and process quantum information. Additionally, they can be efficiently controlled using electromagnetic pulses in the microwave or radio frequency range \cite{Gaita-Ariño2019,Chiesa2024}. The availability of these additional degrees of freedom has recently led to the suggestion of encoding qubits with embedded QEC within single molecules \cite{Chiesa2020,Mezzadri2024,Lim2025}. 
From a chemical point of view, the easiest way to implement this concept is provided by a spin-$S$ ion, characterized by $2S+1$ states. Similarly, giant-spin nanomagnets offer an analogous configuration, as their low-energy spectrum effectively maps onto an effective spin-$S$ system.
Increasing the value of $S$ expands the number of energy levels that can be exploited. 

However, this expansion in computational space comes with greater challenges in controlling decoherence. As the system's dimensionality grows, interactions with the surrounding nuclear bath lead to greater susceptibility to noise and errors. \\
In recent years, significant efforts have been devoted to understanding and mitigating decoherence effects in MNMs \cite{Graham2017,Wedge2012,Kaminski2014}.  For example, hydrogen-free environments have been designed \cite{Santanni2023} to eliminate the effects of fluctuating proton spins, leading to remarkable improvements in coherence times \cite{Zadrozny2015}. Another complementary strategy is to optimize the choice of spin eigenstates encoding quantum information, making them as resilient as possible to the effects of the surrounding bath. Ideally, if these states spanned a decoherence-free subspace (DFS) \cite{Lidar2014} they would be inherently immune to decoherence effects. Yet, a generic bath will dipolarly couple to the full set of molecular spin operators ($3N$, $N$ being the number of spins), preventing a strict realization of DFS conditions. Nevertheless, when the eigenstates are well-separated relative to the dipolar interaction scale, significant opportunities for decoherence mitigation remain. For example, a pair of qubit states in a clock-transition condition has decoherence suppressed to first-order in the qubit-bath interaction. This clock condition is achieved by carefully selecting states with specific symmetries or properties (e.g., anticrossings in the field-dependence of molecular states \cite{Collett2020}).\\
It is important to note that while for a qubit the long $T_2$ time refers to a single, specific pair of eigenstates, a practical implementation of a $d$-dimensional molecular qudit requires long $T_2$ times for superpositions of all $\binom{d}{2}$ pairs of eigenstates. This makes it significantly more challenging to identify optimal setups for mitigating decoherence.
In this paper, we analyze the characteristics of decoherence using a realistic description of molecular qudits and the bath, whose non-Markovian dynamics is accurately computed using a cluster correlation expansion.\\
In section \ref{sec:theory} we show how to model the molecular systems of interest through a spin Hamiltonian approach in a perturbation theory framework. In section \ref{sec:methods} we introduce the technique used to compute the dynamics of the spins.
In section \ref{sec:theoretical_results} we develop a microscopic model of decoherence for molecular spin systems undergoing pure dephasing under a control sequence of $\pi$ pulses, including the Hahn echo \cite{Hahn1950} and the Carr-Purcell-Meiboom-Gill sequence \cite{Carr_Purcell1954,Meiboom_Gill1958}. 
We demonstrate a necessary and sufficient condition to prevent the decay of coherence with time. Moreover, we introduce a parameter to quantify the deviation from this ideal condition. We show that for a coherent superposition of a pair of eigenstates, the key factor driving decoherence is the difference in the expectation values of \textit{local} spin operators between the two states, and not of \textit{total} ones, as one could guess. 
Thus, minimizing these differences can enhance the coherence time of the superposition.

In Section \ref{sec:num_results}, we illustrate this concept using two paradigmatic systems. The first one is a single giant spin $S=10$ highlighting the impact that even tiny differences in the expectation values of spin operators have on the decay of coherence. 
The second one consists of a composite system containing five spins $S_i=1/2$ $(i=1,\dots,5)$, with competing, antiferromagnetic (AFM) interactions. 
It allows us to show the importance of the expectation values of local spin operators over total spin operators in qudits made of several magnetic centers.  
By exploiting the knowledge gained through the aforementioned cases, and accordingly with the experimental constraints, we advance the proposal for a qudit that leverages the properties of the introduced systems.
These results allow us to identify which are the ingredients that can be controlled to engineer optimal molecular qudits for different applications in quantum technologies. Finally, our method is completely general and can be applied to other spin-based quantum systems.

\section{Theoretical background}
\label{sec:theory} 
\subsection{Model Hamiltonian} \label{subsec:ModelHamiltonian}
The spin Hamiltonian describing the problem of a central system of spins interacting with a spin bath can be written as follows:
\begin{equation}
    H = H_S + H_B + H_{SB}
    \label{eq:Htot}
\end{equation}
where $H_S$ is the Hamiltonian of the central system, $H_B$ describes the intrinsic bath interactions, and $H_{SB}$ accounts for the interactions between each element belonging to the central system and the spins contained in the bath.
The central system Hamiltonian has the form
\begin{equation}
H_S=\sum_i\bm{B}\cdot\bm{\Gamma}^i\cdot\bm{S}_i + \sum_{i,j}\bm{S}_i\cdot\bm{D}^{ij}\cdot\bm{S}_j
\label{eq:Hs}
\end{equation}
where $\bm{S}_i$ is the $i^{th}$ system spin coupled to the external magnetic field $\mathbf{B}$ through the gyromagnetic ratio tensor $\mathbf{\Gamma}^i$, whereas $\mathbf{D}^{ij}$ is the tensor describing the interactions between the spin pair $(i,j)$ within the system. If $i=j$,  $\mathbf{D}$ denotes the self-interaction tensor, i.e. the zero field splitting tensor or the quadrupole tensor in the case of electronic and nuclear spins, respectively.
The intrinsic Hamiltonian of the bath is
\begin{equation}
    H_B=\sum_i \bm{B}\cdot\bm{\gamma}^i\cdot\bm{I}_i
    \label{eq:Hb} + \sum_{i,j}\bm{I}_i\cdot \bm{J}^{ij}\cdot\bm{I}_j 
\end{equation}
Similarly to the previous case, $\bm{I}_i$ represents the $i^{th}$ bath spin coupled to the external magnetic field $\mathbf{B}$ through the gyromagnetic ratio tensor $\bm{\gamma}^i$, whereas $\bm{J}^{ij}$ is the interaction tensor bath spins belonging to the pair $(i,j)$.  The case $i=j$ describes the quadrupolar interaction. If $i\ne j$, we assume the two nuclei interacting via magnetic dipole-dipole interaction. In the point dipole approximation, it reads 
\begin{equation}
    \bm{J}^{ij}=\frac{\mu_0}{4\pi\abs{\bm{R}_{ij}}^3}\big(\bm{\gamma}^i\cdot \bm{\gamma}^j-3\frac{\bm{\gamma}^i\cdot \bm{R}_{ij}\otimes \bm{\gamma}^j\cdot\bm{R}_{ij}}{\abs{\bm{R}_{ij}}^2}\big) \qquad i\ne j
    \label{eq:J}
\end{equation}
where $\mu_0$ is the magnetic permeability of free space, $\hbar$ is the reduced Planck's constant, and  $\bm{R}_{ij}$ is the distance vector between the involved spins $i$ and $j$ located respectively in position $\bm{R}_i$ and $\bm{R}_j$.
The system-bath Hamiltonian $H_{SB}$ can be expressed as
\begin{equation}
    H_{SB}=\sum_{i,j} \bm{S}_i\cdot \bm{A}^{ij}\cdot \bm{I}_{j}
    \label{eq:Hsb}
\end{equation}
where $\bm{A}^{ij}$ denotes the interaction tensor between the $i^{th}$ spin belonging to the central system and the $j^{th}$ spin contained in the bath. Once again, a dipole-dipole interaction is assumed:
\begin{equation}
    \bm{A}_{ij}=\frac{\mu_0}{4\pi\abs{\bm{R}_{ij}}^3}\big(\bm{\Gamma}^i\cdot \bm{\gamma}^j-3\frac{\bm{\Gamma}^i\cdot\bm{R}_{ij}\otimes \bm{\gamma}^j\cdot\bm{R}_{ij}}{\abs{\bm{R}_{ij}}^2}\big)
    \label{eq:A}
\end{equation}
with $\bm{R}_{ij}$ being the vector connecting the system spin $i$ and the bath spin $j$. It is worth to note that the methods and results discussed in this paper can be straightforwardly extended to a more general form of $\bm{A}^{ij}$.

In this framework, the total Hamiltonian \ref{eq:Htot} is defined in the Hilbert space $\mathscr{H}=\mathscr{H}_S\otimes\mathscr{H}_B$, where $\mathscr{H}_S$ is the Hilbert space associated with the system, and $\mathscr{H}_B$ the one related to the bath.
A basis set for $\mathscr{H}$ is given by the kets $\ket{\psi\phi}\equiv \ket{\psi}\otimes\ket{\phi}$, where $\ket{\psi}$ and $\ket{\phi}$ are the eigenstates of $H_S$ and $H_B$, respectively.
Labeling the eigenvalues of $H_S$ as $E_S^{\psi}$, the total Hamiltonian may be put in matrix form:

   \begin{equation}
       \begin{aligned}
           \bra{\psi ' \phi '} H \ket{\psi \phi} &= \delta_{\psi ' \psi} \delta_{\phi ' \phi} E_S^{\psi} + \\
       & +\delta_{\psi ' \psi} \bra{\psi ' \phi '} H_B + H_{SB} \ket{\psi \phi} + \\
       & +(1-\delta_{\psi ' \psi}) \bra{\psi ' \phi '} H_{SB} \ket{\psi \phi}
        \end{aligned}
    \end{equation}

where the first term is diagonal in $\mathscr{H}$, the second one collects all the terms of the system-bath coupling that are diagonal in $\mathscr{H}_S$ together with the ones of the bath Hamiltonian. The third term involves states with different $\psi$ in the $H_{SB}$ interaction (off-diagonal elements), which make the treatment of the problem more troublesome. To address this, we apply a canonical transformation to reduce the magnitude of the matrix elements of $H_{SB}$ that couple states with different $\psi$. 
Given that the systems of interest have $H_S$ as the dominant component of the Hamiltonian compared to $H_{SB}$, we adopt a perturbative approach, assuming the dipolar coupling $\bm{A}$ as perturbative parameter  (see equation \ref{eq:Hsb}). 
The application of the canonical, perturbative Schrieffer-Wolff transformation \cite{Slichter1990,Schrieffer1966,Bravyi2011}, yields a unitary transformed Hamiltonian where the off-diagonal terms that are much smaller than the original ones. By neglecting these elements, we finally obtain the following effective Hamiltonian for the bath, conditioned on the system being in the state $\ket{\psi}$:
\begin{equation} 
   H^{\psi}=E_S^{\psi}\mathbb{1}+H_B+H_{SB1}^{\psi}+H_{SB2}^{\psi}
   \label{eq:SWH}
\end{equation}
$\mathbb{1}$ being the identity operator in the {$\mathscr{H}_B$ Hilbert space. $H_{SB1}$ and $H_{SB2}$
are the central system-bath interaction terms treated at first and second order in perturbation theory, respectively. They have the following form:
\begin{equation}
    H_{SB1}^{\psi}=\sum_{i,j} \bra{\psi}\bm{S}_i\ket{\psi}\cdot \bm{A}^{ij}\cdot \bm{I}_j
    \label{eq:Hsb1}
\end{equation}
and
\begin{equation}
\begin{split}
    & H_{SB2}^{\psi}=\sum_{j,l} \bm{I}_j\cdot \bm{T}^{jl}(\psi)\cdot \bm{I}_l \qquad\\ \text{where} \\
    & \bm{T}^{jl}(\psi)=\sum_{\psi ',i,k} \frac{\bra{\psi}\bm{S}_i\ket{\psi ' }\cdot \bm{A}^{ij}\otimes \bra{\psi '}\bm{S}_k\ket{\psi}\cdot \bm{A}^{kl}}{E_{\psi}-E_{\psi ' }} \\
    \label{eq:T}
\end{split}
\end{equation}
It is important to stress that both \ref{eq:Hsb1} and \ref{eq:T} depend on the specific state $\ket{\psi}$ of the central system. This aspect will be crucial in the description of the temporal dynamics of a coherent superposition of central system states, as it will be extensively discussed in the following.

\subsection{Decoherence}
\label{subsec:Decoherence}
At low temperatures ($\approx 1 K$ or below), the main source of decoherence for MNMs is represented by the coupling of the central system with the surrounding spin bath. Indeed, in this regime the contribution to decoherence due to the interactions with phonons is usually negligible. Moreover, many experiments on molecular spin complexes have shown that the relaxation time $T_1$ becomes several orders of magnitude longer than the dephasing time \cite{Fataftah2016,Ding2016,Bader2016}. Therefore, within the timescale we are interested in, the state populations (i.e. the diagonal terms of its density matrix) result unaffected.
Here, we want to model how the generic superposition between the two eigenstates $\ket{\alpha}$ and $\ket{\beta}$ of equation \ref{eq:Hs} is affected by pure-dephasing due to the system-bath interaction.
Thus we consider the initial density matrix for the overall system (central system + bath) $
    \ket{\psi}_{\alpha, \beta}\bra{\psi}_{\alpha, \beta}\otimes \rho_B$, where $\rho_B$ is the initial density matrix of the bath and
\begin{equation}
    \ket{\psi}_{\alpha, \beta}=C_{\alpha}\ket{\alpha}+C_{\beta}\ket{\beta}
    \label{eq:system_state}
\end{equation}
is the initial state of the system with coefficients  $C_{\alpha}$ and $C_{\beta}$.
The whole density matrix evolves in time according to the Von-Neumann equation through the Hamiltonian $\bigoplus_{\psi}H^{\psi}$, block diagonal in $\mathscr{H}$ and with $H^{\psi}$ defined in equation \ref{eq:SWH}. 
Due to the entanglement between the system and the bath, after a time $\tau$, the coherence term of the reduced density matrix for the system will be
\begin{equation}
    C_{\alpha}C_{\beta}^* \,tr_B\big(e^{-iH^{\alpha}t}\rho_B e^{iH^{\beta}t} \big)
\end{equation}
where $t=\tau/ \hbar$. 
Note that this is the only non-trivial coherence term in the reduced density matrix because we considered a superposition of two eigenstates.
Now, we define the coherence factor, as the ratio between the final and the initial coherence, that is
\begin{equation}
    L^{\alpha\beta}(t) = tr_B\big(e^{iH^{\beta}t}e^{-iH^{\alpha}t}\rho_B \big)
    \label{eq:coherence}
\end{equation}
with the initial condition $L^{\alpha\beta}(0)=1$ because, at $t=0$, the coherence is $C_{\alpha}C_{\beta}^*$. 
To recover the central system coherence factor loss over time, a powerful approach known as dynamical decoupling was developed \cite{Viola1998,Viola1999}. The key idea is to effectively decouple the central spin from the environment by frequently flipping the central system spin through a series of electromagnetic pulses. Indeed, this procedure allows us to switch $H^{\alpha}$ and $H^{\beta}$ during the time propagation, partially compensating the pure dephasing induced by these Hamiltonians.
Therefore, equation \ref{eq:coherence} can be generalized by supposing the application of a set of instantaneous pulses on the system that swaps the probability amplitude of the states $\alpha$ and $\beta$.
The time evolution operators related to the application of $k$ pulses are
\begin{equation}
    U^{\alpha(\beta)}(t)=
    \begin{cases}
    \begin{aligned}
       & e^{-iH^{\alpha(\beta)}\Delta t_1} \qquad &k=0 \\
    & \prod_{j=1}^k e^{-iH^{\beta(\alpha)} \Delta t_{j}}e^{-iH^{\alpha(\beta)} \Delta t_{j}} \qquad &k>0 \\ 
    \end{aligned}   
    \end{cases}    
    \label{eq:propagator}
\end{equation}
where $\Delta t_j=t_j-2\sum_{l=1}^j\Delta t_{l-1}$, with $\Delta t_0=0$ and $t\in \mathbb{R}^k$, specifies the free evolution time before and after the $j^{th}$ pulse $\forall j>1$. 
From a physical viewpoint, if $k=0$ (i.e. no electromagnetic pulses are applied), the system undergoes a free induction decay, whereas $k=1$ implies the use of the Hahn echo technique and, for $k>1$, the CPMG pulse sequence is represented.
In this general framework, the coherence factor assumes the following form
\begin{equation}
    L^{\alpha\beta}(t) = tr_B\big(U^{\beta}(t)^{\dag}U^{\alpha}(t) \rho_B \big)
    \label{eq:coherence_pulses}
\end{equation}
and, its square modulus, is the quantity we want to preserve.
Note that we did not assume neither that $\rho_B$ is thermal at each time, nor a Markovian dynamics for the system, thus the resulting process cannot be described in terms of Lindblad master equations \cite{Breuer2007,Manzano2020}.

\section{\label{sec:methods} Methods}

To calculate the coherence factor defined in equation \ref{eq:coherence_pulses}, it is necessary to evaluate the time propagators for the bath. However, even for small baths, it is impossible to numerically carry out this task due to the exponential growth of the computational costs. To this end, we resorted to the Cluster Correlation Expansion (CCE) approach \cite{Yang2008,Yang2009,Yang2017} which allows one to approximate the coherence factor with its product expansion truncated at a certain order. More specifically, this technique represents a systematic method to take into account the many-body correlations in the bath, order by order.
By considering a Hamiltonian diagonal in the system Hilbert space, like the one in equation \ref{eq:SWH}, we define first of all the coherence factor given by a certain cluster $C$ belonging to the bath as
\begin{equation}
    L_C^{\alpha \beta}(t)=tr\big(U_C^{\beta}(t)^{\dag}U_C^{\alpha}(t) \rho_B \big)
\end{equation}
where $U_C^{\alpha (\beta)}(t)$ is the propagator for the cluster $C$. This is obtained by substituting the Hamiltonian $H^{\alpha (\beta)}$ in the exponential of equation \ref{eq:propagator} with the one associated with the cluster $C$, so replacing all the operator $\bm{I}_i$ in $H^{\alpha (\beta)}$ $\forall i\notin C$ with their mean-field average.
The coherence factor \ref{eq:coherence_pulses} can be expanded as
\begin{equation}
    L^{\alpha \beta}(t)=\prod_C \Tilde{L}_C^{\alpha \beta}(t)
    \label{eq:cce}
\end{equation}
where $C$ runs over all the possible clusters contained in the spin bath. The cluster correlations are recursively defined as
\begin{equation}
   \Tilde{L}_C^{\alpha \beta}(t)=\frac{L_C^{\alpha\beta}(t)}{\prod_{\Tilde{c}}\Tilde{L}_{\Tilde{c}}^{\alpha\beta}(t)} 
\end{equation}
where $\Tilde{c} \subset C$. 
For example, to evaluate the coherence factor term
\begin{equation}
   \Tilde{L}_{\{i,j\}}^{\alpha\beta}(t)=\frac{L_{\{i,j\}}^{\alpha\beta}(t)}{\Tilde{L}_0^{\alpha\beta}(t)\Tilde{L}_{\{i\}}^{\alpha\beta}(t)\Tilde{L}_{\{j\}}^{\alpha\beta}(t)}
\end{equation}
associated to the cluster $\{i,j\}$, where $i,j$ are two bath spins, we need to evaluate the following terms of the expansion
\begin{equation}
    \begin{split}
        & \Tilde{L}_0^{\alpha\beta}(t)=L_0^{\alpha\beta}(t) \in \mathbb{C} \\
        & \Tilde{L}_{\{i\}}^{\alpha\beta}(t)=L_{\{i\}}^{\alpha\beta}(t)/\Tilde{L}_0^{\alpha\beta}(t) \\
        & \Tilde{L}_{\{j\}}^{\alpha\beta}(t)=L_{\{j\}}^{\alpha\beta}(t)/\Tilde{L}_0^{\alpha\beta}(t) \\
    \end{split}
\end{equation}

The CCE theory is related to the Linked Cluster Expansion (LCE) \cite{Abrikosov2012,Mahan2000}. Indeed, the LCE approach for qubit decoherence has been developed for bath spins with $S=1/2$ and $S>1/2$ respectively in \cite{Saikin1017} and \cite{Yang2008}, showing that
\begin{equation}
    \Tilde{L}_C=e^{\pi (C)}
\end{equation}
where $\pi (C)$ is the sum of all the connected Feynman diagrams in which all and only the spins in $C$ underwent a flip-flop transition, demonstrating that $\Tilde{L}_C$ is a summation of infinite diagrams.

For all the molecular systems considered below, the expansion in equation \ref{eq:cce} has been truncated to its second order (2-CCE),
\begin{equation}
    L^{\alpha \beta}(t)\approx \Tilde{L}_0^{\alpha\beta}(t) \prod_i \Tilde{L}_{\{i\}}^{\alpha\beta}(t) \prod_{i,j} \Tilde{L}_{\{i,j\}}^{\alpha\beta}(t) 
\end{equation}
Indeed, within the timescale considered, we have verified that spin clusters containing two elements are sufficient to cause complete decoherence, leading to convergence in the CCE. This order of convergence in the expansion is in agreement with other results in literature where the CCE method is applied to molecular spin systems \cite{Troiani2008,Petiziol2021}.
An open-source Python package \cite{Onizhuk2021} that implements the CCE theory for a central system of spins undergoing environment induced decoherence is already available. Nevertheless, in order to meet the specific needs of our study, we developed from scratch a home made code.

Equipped with a model to describe decoherence and a method to properly simulate the behavior of a many-body bath, we proceed by defining a strategy to counteract decoherence itself.

\section{Theoretical results}
\label{sec:theoretical_results}
In the present section, we will prove a necessary and sufficient condition on the Hamiltonian operators $H^{\alpha}$ and $H^{\beta}$ to keep the coherence factor $L^{\alpha\beta}(t)$ equal to 1. Moreover, we will show with analytical arguments that this condition can be expressed in terms of differences in expectation values of
its local spin operators.
\subsection{Hamiltonian and decoherence}
\label{sub:Hamiltonian_and_decoherence}
In this subsection, we will prove that $L^{\alpha\beta}(t)=1$  for each bath reduced density matrix $\rho_B$, $\forall k\ge 1$ and $\forall t$ if and only if
\begin{equation}
    [H^{\alpha},H^{\beta}]=0
    \label{eq:commutator}
\end{equation}

First of all, we will prove the \textit{only if} implication.
We start by writing the condition $L^{\alpha\beta}(t)=1$ as
\begin{equation}
    tr_B\big(\big[ U^{\beta}(t)^{\dag}U^{\alpha}(t)-\mathbb{1}\big] \rho_B \big)=0
\end{equation}
where we used the fact that $tr_B(\rho_B)=1$.
Since the Hilbert-Schmidt inner product between the term in the square brackets and $\rho_B$ must be $0$ for each $\rho_B$, we find that
\begin{equation}
    U^{\beta}(t)^{\dag}U^{\alpha}(t)=\mathbb{1}
\end{equation}
that can be written as
\begin{equation}
    \prod_{l=1}^{k}e^{iH^{\beta}\Delta t_{k-l+1}}e^{iH^{\alpha} \Delta t_{k-l+1}}\prod_{m=1}^{k}e^{-iH^{\beta} \Delta t_{m}}e^{-iH^{\alpha}\Delta t_{m}}=\mathbb{1}
\label{eq:order_k}
\end{equation}

Because of the independence of the $\Delta t_l$ variables, we can write
\begin{widetext}
    \begin{equation}
    \begin{aligned}
     M&=e^{-iH^{\alpha}\Delta t_{1}}e^{-iH^{\beta} \Delta t_{1}}e^{iH^{\alpha}\Delta t_{1}}e^{iH^{\beta} \Delta t_{1}}= 
e^{-i\Delta t_{1}(H^{\alpha}+H^{\beta})-\frac{\Delta t_{1}^2}{2}[H^{\alpha},H^{\beta}]+...}e^{+i\Delta t_{1}(H^{\alpha}+H^{\beta})-\frac{\Delta t_{1}^2}{2}[H^{\alpha},H^{\beta}]+...} = \\
    &= e^{\sum_{n=2}^{\infty} \Delta t_1^n f_n(H^{\alpha}, H^{\beta})} \\
    \label{eq:M}
\end{aligned}
\end{equation}
\end{widetext}

where the matrix $M$ does not depend on any variable and $f_n(H^{\alpha}, H^{\beta})$ are functions of the Hamiltonians obtained using the well-known Baker – Campbell – Hausdorff formula. Since $M$ is constant for each $\Delta t_1\in \mathbb{R}$, we must have that $f_n(H^{\alpha}, H^{\beta})=0 \quad \forall n$, thus $M=\mathbb{1}$.
For $n=2$ we note that $f_2(H^{\alpha}, H^{\beta})=-\Delta t_{1}^2[H^{\alpha},H^{\beta}]$, then this condition is fulfilled only if $[H^{\alpha},H^{\beta}]=0$. Since, by definition, we know that also the other terms $f_n(H^{\alpha}, H^{\beta})$ are zero if $[H^{\alpha},H^{\beta}]=0$, then we demonstrated our claim. 
Now the \textit{if} implication can be easily proved. Indeed, by supposing that $[H^{\alpha},H^{\beta}]=0$ is true, from equation \ref{eq:M} we get that $M=1$ because $f_n(H^{\alpha}, H^{\beta})=0 \quad \forall n$, as outlined above. Thus we satisfied equation \ref{eq:order_k}, obtaining that $L^{\alpha\beta}(t)=1$.

Thus we showed the bijective relation between the coherence factor and the commutation of the Hamiltonian operators.
Note that, since $M=\mathbb{1}$, the coherence factor is preserved after each pulse $j=1,\dots,k$ because, as we assumed in equation \ref{eq:propagator}, the free propagation before and after the pulse is the same.

\subsection{Decoherence and expectation values of local spin operators}
\label{subsec:Decoherence&SpinTexture}
The content of this subsection is dedicated to the demonstration of the connection between the coherence factor and the expectation values of its local spin operators starting from equation \ref{eq:commutator}. 
By taking into account the Hamiltonian in equation \ref{eq:SWH}, we can write the commutator of two Hamiltonian operators, conditioned on the states $\ket{\alpha}$ and $\ket{\beta}$, as
\begin{equation}
\begin{aligned}
[H^{\alpha},H^{\beta}]&=[H_{SB1}^{\alpha},H_{SB1}^{\beta}]+[H_B,\Delta H_{SB1}]+\\
    & +[H_B,\Delta H_{SB2}]+[H_{SB1}^{\alpha},H_{SB2}^{\beta}]+\\
    & +[H_{SB2}^{\alpha},H_{SB1}^{\beta}]+[H_{SB2}^{\alpha},H_{SB2}^{\beta}] \\
\end{aligned}
\label{eq:full_commutator}
\end{equation}

where $\Delta H_{SB1(2)}=H_{SB1(2)}^{\beta}-H_{SB1(2)}^{\alpha}$.
The first term on the right can be written as
\begin{widetext}
    \begin{equation}
\begin{split}
    [H_{SB1}^{\alpha},H_{SB1}^{\beta}] &=\sum_{\substack{k,l,j \\ \mu, \nu,\zeta,\delta,\theta}} i\bra{\beta}S_k^{\mu}\ket{\beta}\bra{\alpha}S_l^{\zeta}\ket{\alpha}A_{\mu\nu}^{kj}A_{\zeta\delta}^{lj}\epsilon_{\nu\delta\theta}I_j^{\theta} = \\
    & =\sum_{\substack{k,l,j \\\mu, \nu,\zeta,\delta,\theta}} \frac{i}{2}\Bigl(\bra{\beta}S_k^{\mu}\ket{\beta}-\bra{\alpha}S_k^{\mu}\ket{\alpha}\Bigr)\Bigl(\bra{\beta}S_l^{\zeta}\ket{\beta}+\bra{\alpha}S_l^{\zeta}\ket{\alpha}\Bigr) A_{\mu\nu}^{kj}A_{\zeta\delta}^{lj}\epsilon_{\nu\delta\theta}I_j^{\theta}
\end{split}
\end{equation}
\end{widetext}
where $k$, $l$ span over the central system spins, $j$ refers to a bath spin, whereas $\mu,\zeta,\nu,\delta,\theta=x,y,z$ and $\epsilon_{\nu\delta\theta}$ is the Levi-Civita tensor.
For the second one we get
\begin{equation}
    [H_B,\Delta H_{SB1}]=\sum_{\substack{k,j\\\mu,\nu}} \big(\bra{\beta}S_k^{\mu}\ket{\beta}-\bra{\alpha}S_k^{\mu}\ket{\alpha}\big)A_{\mu\nu}^{kj}[H_B,I_j^{\nu}]
\end{equation}
and equation \ref{eq:full_commutator} becomes
\begin{widetext}
    \begin{equation}
    \begin{split}
        [H^{\alpha},H^{\beta}]= \sum_{\substack{k,l,j,\\ \mu, \eta,\nu,\zeta,\delta,\theta}} \Bigl(\bra{\beta}S_k^{\mu}\ket{\beta}-\bra{\alpha}S_k^{\mu}\ket{\alpha}\Bigl)\Gamma_{\eta\mu}^k 
        \Tilde{A}_{\eta\nu}^{kj}\biggl(\Bigl(\bra{\beta}S_l^{\zeta}\ket{\beta}+\bra{\alpha}S_l^{\zeta}\ket{\alpha}\Bigl)\frac{i}{2}A_{\zeta\delta}^{lj}\epsilon_{\nu\delta\theta}I_j^{\theta}+[H_B,I_j^{\nu}]\biggl) + \Tilde{R}
    \end{split} 
    \label{eq:expanded_comm}
\end{equation}
\end{widetext}
where, again, $k,l,j$ refer to specific spins, $\mu,\eta,\zeta,\nu,\delta,\theta=x,y,z$  and
\begin{equation}
    \Tilde{\bm{A}}^{kj}=\frac{\mu_0}{4\pi\abs{\bm{R}_{kj}}^3}\big(\bm{\gamma}^j-3\frac{\bm{R}_{kj}\otimes \bm{\gamma}^j\cdot \bm{R}_{kj}}{\abs{\bm{R}_{kj}}^2}\big)
    \label{eq:B}
\end{equation}
In equation \ref{eq:expanded_comm}, the operator $\Tilde{R}$ includes all the commutators, appearing in equation \ref{eq:full_commutator}, containing the terms $H_{SB2}^{\alpha(\beta)}$ at least once.
In the supplementary information section (see \ref{subsec:suppl_neglect}), we will show that, under the proper hypothesis, $\Tilde{R}$ can be neglected. If that is the case, we get 
\begin{equation}
    [H^{\alpha},H^{\beta}] \approx \sum_{\substack{k,\mu,\eta}} (\bra{\beta}S_k^{\mu}\ket{\beta}-\bra{\alpha}S_k^{\mu}\ket{\alpha})\Gamma_{\eta\mu}^k O_{\eta}^k
 \label{eq:approx_commutator}
\end{equation}
where the explicit form of the $O_{\eta}^k$ operators can be derived from equation \ref{eq:expanded_comm}. Note that these operators contain all the information involving the nuclear spin bath parameters, including the relative distances between the system and bath spins, which cannot be controlled since we are outlining a strategy to engineer the system only. Equation \ref{eq:approx_commutator} allows us to isolate and highlight the key ingredients affecting the coherence dynamics that depend uniquely on the central system. They all appear in the form of differences between expectation values of \textit{local} spin operators, which are strictly related to the structure of the eigenstates $\ket{\alpha}$ and $\ket{\beta}$. By acting on them, we aim to reduce the value of the commutator between the Hamiltonians to approach condition \ref{eq:commutator} and, therefore, to increase the correlation factor.
To this purpose, we define $W_k=\{i:\bm{R}_i \equiv \bm{R}_k \}$, the equivalence class of spins of the system that share the same spatial position of the spin $k$. Indeed, each class can contain up to two elements in the case of a nuclear spin and the electron associated with it. 
It follows that equation \ref{eq:approx_commutator} satisfies relation \ref{eq:commutator} if 
\begin{equation}
   \sum_{\substack{i\in W_k\\ \mu}} \bra{\beta}S_i^{\mu}\ket{\beta}\Gamma_{\eta\mu}^i=\sum_{\substack{i\in W_k\\ \mu}} \bra{\alpha}S_i^{\mu}\ket{\alpha}\Gamma_{\eta\mu}^i \qquad \forall \eta,W_k
   \label{eq:general_spin_texture}
\end{equation}
From a physical viewpoint this means that, if the expectation values of the spin operators on $\ket{\alpha}$ and $\ket{\beta}$ are the same for each spatial direction $\eta$ and site equivalence class $W_k$, the evolution in time of the bath under the two Hamiltonians generated by the pulses is the same. 
If $\Gamma_{\eta\mu}^i=\delta_{\eta\mu}\Gamma_{\eta}^i$ and the system spins are located in different spatial positions, we get the following equation, which offers a clearer physical interpretation:  
\begin{equation}
    \bra{\beta}S_k^{\eta}\ket{\beta}=\bra{\alpha}S_k^{\eta}\ket{\alpha} \qquad \forall k, \eta
   \label{eq:spin_texture}
\end{equation}
Thus, in this case, condition \ref{eq:commutator} is fulfilled if the expectation values of the \textit{local} spin operators, when evaluated over the two system eigenstates combined in the initial coherent superposition reported in equation \ref{eq:system_state}, are exactly the same. In principle, all the spatial components of the spin operators $S_k$ should be considered. In practice, in systems like the ones that we will show in the following, where the leading interaction is represented by the Zeeman term (first term at second member in equation \ref{eq:Hs}), only the component ($z$)  parallel to the external field is relevant.       
The closer we are to the ideal conditions expressed in equations \ref{eq:general_spin_texture} or \ref{eq:spin_texture}, the more the bath cannot distinguish in which order the propagators associated with $H^{\alpha}$ and $H^{\beta}$ are applied. As a result, the Hilbert space generated by $\ket{\alpha}$ and $\ket{\beta}$ effectively behaves as a decoherence-free (DF) subspace \cite{Zanardi1997}, in agreement with the specific result shown in \cite{Troiani2012}.
Moreover, since $\ket{\alpha}$ and $\ket{\beta}$ are eigenstates, the decoherence cannot emerge from the dynamics due to the system Hamiltonian \cite{Lidar1998}. 
In the case of qubits, one possible way to satisfy condition \ref{eq:spin_texture} is by encoding information in two eigenstates that undergo an anticrossing at a specific applied field, thereby realizing clock-transition conditions \cite{Collett2020}. At this anticrossing, the transition frequency between the two levels becomes first-order insensitive to small fluctuations in the magnetic field, and both levels exhibit equal magnetization:
\begin{equation}
    \sum_k \Gamma_z^k \bra{\beta}S_k^z\ket{\beta}=\sum_k \Gamma_z^k \bra{\alpha}S_k^z\ket{\alpha} 
   \label{eq:clock}
\end{equation}
This condition can imply \ref{eq:spin_texture}, for instance, through symmetry arguments, but not necessarily. The most favorable condition is when the clock transition occurs at zero applied field, as in this case time-reversal symmetry enforces $\bra{\alpha}S_k^{z}\ket{\alpha}=\bra{\beta}S_k^{z}\ket{\beta}=0$ $\forall k$.\\ For qudits, the challenge lies in the relevance of multiple transition frequencies, making it extremely difficult to satisfy condition \ref{eq:clock} for all of them—except at zero applied field. This would require a non-Kramers nanomagnet (with an even number of electrons) that possesses sufficiently low symmetry, ensuring the presence of $d$ low-lying nonmagnetic singlet eigenstates. These states must exhibit well-spaced energy gaps of similar magnitude to remain within an addressable energy window compatible with available magnetic pulses. Additionally, effective pulse operations require sizeable magnetic-dipole matrix elements between multiple pairs of states. Finally, no energy gap should be so small that second-order decoherence contributions become significant.\\Meeting all these constraints in a single physical system is highly challenging, particularly because the zero-field requirement precludes any external control over the Hamiltonian. Therefore, rather than aiming for a perfectly optimal control of decoherence, we will explore whether suboptimal but still viable conditions can be achieved in realistic systems and experimental setups, and whether for such systems the effect of second-order terms in decoherence is actually minor.
We stress that we do not require Markovian dynamics \cite{Chiesa2022} because we only need the applicability of the Schrieffer-Wolff transformation for the eigenstates of interest, extending the range of applicability of our method.

\subsection{Consequences of neglecting the second order perturbed terms}
\label{subsec:neglect_sec_order}
In section \ref{subsec:Decoherence&SpinTexture}, we stated that the term $\Tilde{R}$ in equation \ref{eq:expanded_comm} can be neglected. In supplemental material \ref{subsec:suppl_neglect}, we demonstrate that for MNMs the second-order perturbed terms can be neglected by estimating all the interactions acting on a generic spin in the bath. The results show that first-order interactions (Equation \ref{eq:Hsb1}) are usually much larger than second-order ones (Equation \ref{eq:T}), unless the terms $(\bra{\beta}S_k^{\mu}\ket{\beta}-\bra{\alpha}S_k^{\mu}\ket{\alpha})\Gamma_{\eta\mu}^k$ in \ref{eq:expanded_comm} are small. In this case (see supplemental material \ref{subsec:suppl_neglect})
\begin{equation}
    \begin{aligned}
    [H^{\alpha},H^{\beta}] \approx &\sum_{\substack{k,\mu,\eta}} (\bra{\beta}S_k^{\mu}\ket{\beta}-\bra{\alpha}S_k^{\mu}\ket{\alpha})\Gamma_{\eta\mu}^k O_{\eta}^k+\\
    &+[H_B,\Delta H_{SB2}].
    \label{eq:approx_comm_2}
\end{aligned}
\end{equation}
Yet, even when it is the leading term in \ref{eq:approx_comm_2}, the second-order contribution is usually small, with little impact on coherence times.

\section{Numerical results}
\label{sec:num_results}
In this section, we illustrate the impact on the coherence factor when the ideal condition expressed in Equation \ref{eq:general_spin_texture} is not met, using two representative case studies. Additionally, we present an example of a realistic molecular spin qudit.

For this purpose, we define the following parameter 
\begin{equation}\label{eq:spin_texture_general}
    \Delta = \sum_{\substack{W_k\\ \eta}}\abs{\sum_{\substack{i\in W_k\\ \mu}} \Gamma_{\eta\mu}^i(\bra{\beta}S_i^{\mu}\ket{\beta}-\bra{\alpha}S_i^{\mu}\ket{\alpha})}
\end{equation}
that measures how far we are from the ideal condition expressed in equation \ref{eq:commutator}.
Note that, in the following, we will study central systems made of electronic spins only. In addition, we suppose diagonal gyromagnetic ratio tensors. It follows that, considering equation \ref{eq:spin_texture}, equation \ref{eq:spin_texture_general} takes the following simplified form
\begin{equation} \label{eq:spin_texture_simple}
    \Delta = \sum_{k,\eta}\abs{ \Gamma_{\eta\eta}^k(\bra{\beta}S_k^{\eta}\ket{\beta}-\bra{\alpha}S_k^{\eta}\ket{\alpha})}
\end{equation}
All the numerical values of $\Delta$ reported in this work will be expressed in Bohr magneton units.
In all the systems we considered, a Hanh echo pulse sequence has been considered to suppress the trivial effect of the free induction decay, modeled as explained in section \ref{subsec:Decoherence}.
Besides, the randomly generated spin baths are composed of 1000 nuclear spins ($I=1/2$) distributed into a spherical volume of radius $20\,\text{\AA}$, with the only constraint of setting the minimum distance between each pair of spins (belonging to the system or the bath) to a typical value of 3 \AA.

\subsection{\label{subsec:singleGiant}Single giant spin-S system}
As a first example, we report the case of the simplest molecular spin qudit, namely a system that contains a single spin $S>1/2$. Specifically, we consider a giant electronic spin with 
$S=10$, which provides a large set of states for potential logical encoding.
Note that there are no real representatives of single ions with an electronic spin $S=10$. However, there are many examples of molecular clusters composed of many strongly interacting metal ions, whose spins are locked at low temperatures into an effective ``giant spin''. In these cases, such as the well known Mn$_{12}$ and Fe$_8$ MNMs \cite{Mirebeau1999, Caciuffo1998}, the giant spin describes the whole magnetic core of the molecule, and within the ground multiplet local spin operators are all proportional to the total spin operators, $\mathbf{S}_k = A_k \mathbf{S}$. Thus, the low-energy Hamiltonian and equation \ref{eq:spin_texture_simple} can be formulated in terms of the total spin only.
Given the illustrative purpose of the present example, to maintain simplicity and avoid specifying the internal spin structure of any particular MNM, we assume $k$-independent values for  $\Gamma_{\eta\eta}^k$ and projection coefficients $A_k$ (i.e., $A_k=S/N_s$, $N_s$ being the number of ions). Thus, Equation \ref{eq:spin_texture_simple} can be expressed solely in terms of the total spin.
The effective Hamiltonian has the form:
\begin{equation} \label{eq:HSingleGiant}
    H_S = \bm{S} \cdot\bm{D}_{ZFS}\cdot \bm{S}  + \bm{B}\cdot \bm{\Gamma} \cdot \bm{S},
\end{equation}
where \textbf{D$_{ZFS}$} is the zero field splitting tensor, \textbf{B} is the external magnetic field vector, and $\bm{\Gamma}$ is the gyromagnetic ratio tensor. \textbf{D$_{ZFS}$} can be written as
\begin{equation*}
    \textbf{D}_{ZFS} = 
    \begin{pmatrix}
        -D/3 + E & 0 & 0 \\
        0 & -D/3 - E & 0 \\
        0 & 0 & 2D/3
    \end{pmatrix},    
\end{equation*}
where $D$ and $E$ are defined starting from the initial components of the tensor $D_{ZFS}^{xx}$,$D_{ZFS}^{yy}$,$D_{ZFS}^{zz}$, referred to the $x,y,z$ coordinate axes: $D = D_{ZFS}^{zz}-\frac{1}{2}D_{ZFS}^{xx}-\frac{1}{2}D_{ZFS}^{yy}$, $E = \frac{1}{2}(D_{ZFS}^{xx}-D_{ZFS}^{yy})$. We assume a realistic value of $D =$ $25$ $\mu eV$ \cite{Mirebeau1999, Caciuffo1998} and a low level of rhombicity, with $E/D=0.02$.
The magnetic field is supposed to be oriented along the $z$ direction, i.e. \textbf{B} $\equiv$ B\bm{$\hat{z}$}, with B = 0.07 T. Moreover, we consider a diagonal, isotropic $\bm{\Gamma}$ tensor, with components $\Gamma_{xx}=\Gamma_{yy}=\Gamma_{zz}=2.0$ $\mu_B$, where $\mu_B$ is the Bohr's magneton.  After diagonalization of the Hamiltonian \ref{eq:HSingleGiant}, the energy levels are labeled from $0$ to $20$ following an ascending order across the spectrum. Correspondingly, the associated eigenstates are labeled from $\ket{0}$ to $\ket{20}$.
Since $E$ is small, the anisotropy nearly represents a double-well potential. Eigenstates can then be associated to one of the two  wells depending on the sign of $\langle S_z \rangle$. Note that the sign of $D$ corresponds to an easy-plane anisotropy, with small values of $\langle S_z \rangle$ in low-lying states. 

\begin{figure}[h]
    \includegraphics[width=0.95\linewidth]{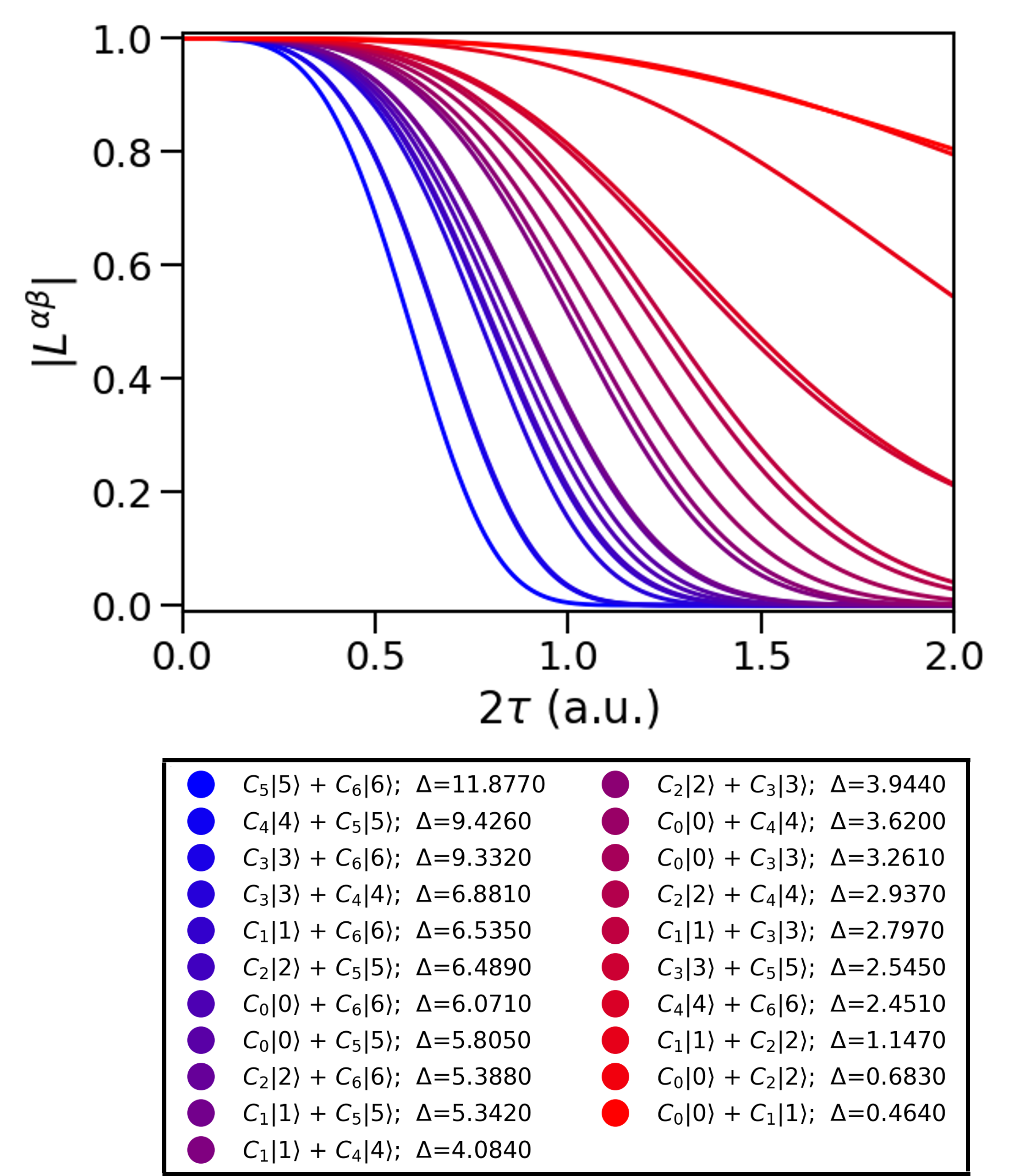}
    \caption{Top: coherence factor over time of the pair superpositions of states corresponding to seven lowest energy levels of the spectrum. Bottom: legend displaying the representative color for each superposition and the corresponding value of $\Delta$.}
    \label{fig:venPairs_decoherence+legend}
\end{figure}
In order to show the link between the coherence factor of a superposition of two eigenstates and the relative $\Delta$ parameter as defined in \ref{eq:spin_texture_simple} and depicted in section \ref{subsec:Decoherence&SpinTexture}, we start by considering the set of seven lowest energy levels within the spectrum. For each possible pair of states, a coherent superposition is prepared.  

The characteristic values of $\Delta$ are calculated for each pair. and are summarized in the legend in figure \ref{fig:venPairs_decoherence+legend}.  
The coherence factor decays, computed by using the CCE method (up to second order), are displayed in figure \ref{fig:venPairs_decoherence+legend}. The timescale is normalized by taking as a reference the time employed by the fastest curve to reach the value of 0.001, which is approximately $0.1\, \mu s$.
It is evident that, by gradually decreasing the value of $\Delta$ with a proper choice of $\alpha$ and $\beta$, it is possible to increase the value of the coherence factor of the system. This result is in perfect agreement with the theoretical calculations carried out in subsections \ref{sub:Hamiltonian_and_decoherence} and \ref{subsec:Decoherence&SpinTexture}. We stress the fact that even differences on the second decimal place of $\Delta$ contribute to the ordering of the curves.   It is important to notice that, in principle, one should evaluate the expectation values of the $x$ and $y$ components of $S$ too, according to what has been stated in equation \ref{eq:spin_texture}. However, in this specific case, those quantities are null $\forall$ $\alpha,\beta$.

What we have shown so far is a good, illustrative example of the concrete implications of the theory developed in section \ref{sec:theory}. The more similar the eigenstates of a superposition from a magnetic point of view (expectation values of spin magnetic moments), the lower the value of $\Delta$ and, therefore, the longer the coherence time. Further insights can be gained by considering composite systems made of multiple spins spatially located in different positions, as it will be discussed in the following section.

\subsection{\label{subsec:5_spins} System with competing interactions}
According to the definition of the parameter $\Delta$ in \ref{eq:spin_texture_general} and \ref{eq:spin_texture_simple}, we expect that the leading terms causing decoherence are differences in the expectation values of \textit{local} spin operators, instead of \textit{total} spin operators, as one could expect at first glance.  
To demonstrate this result, we propose as a second example a system composed of five $S=1/2$ spins interacting through competing interactions, shown in figure \ref{fig:frustrated_system}.

\begin{figure}[h]

\begin{subfigure}{0.6\columnwidth}
\includegraphics[width=\linewidth]{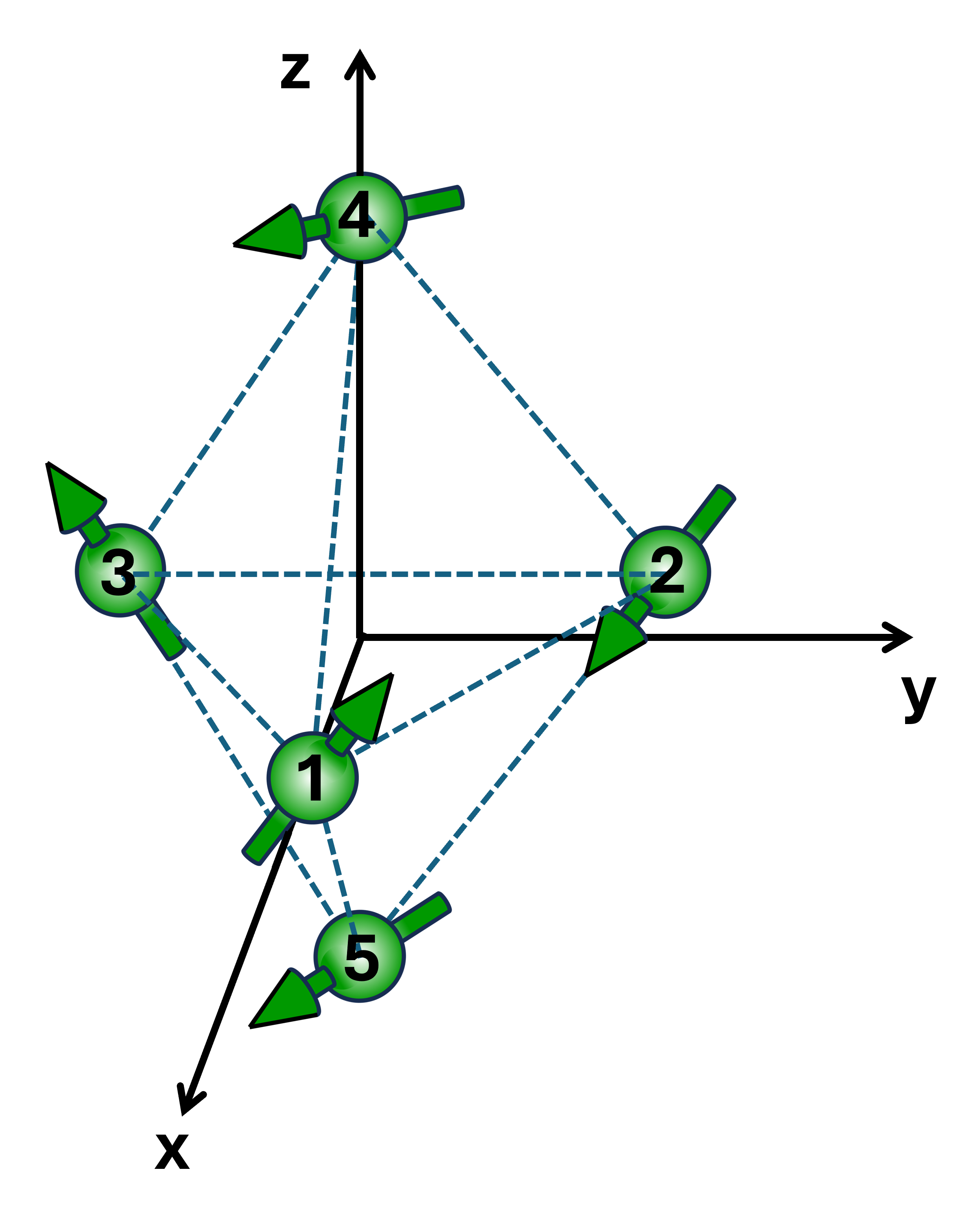}
\caption{}
\label{fig:frustrated_system}
\end{subfigure}
\begin{subfigure}{0.6\columnwidth}
\includegraphics[width=\linewidth]{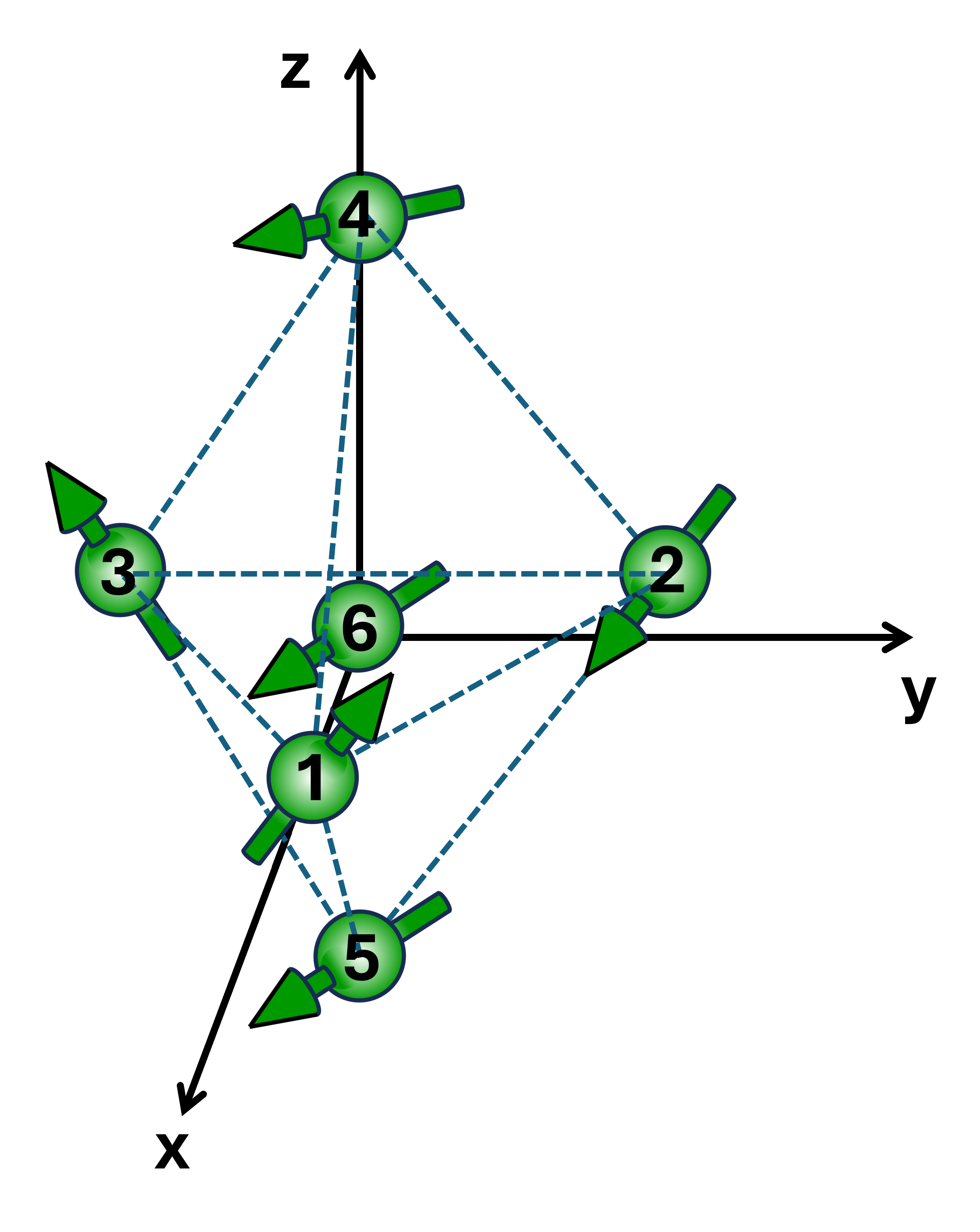}
\caption{}
\label{fig:frustratedQudit}
\end{subfigure}

\caption{(a) Schematic structure of the system made of five spins 1/2. Spins $1,2,3$ are placed at the vertices of an equilateral triangle lying on the $x-y$ plane while spins $4$ and $5$ are placed along the $z-$axis. Each spin is positioned at a distance of 3 \AA  $\,$ from the center. However, even when this symmetry is disrupted, the qualitative results remain unchanged.}
The resulting solid consists of a double tetrahedron.(b) For the qudit proposal, a sixth spin is added at the origin of the reference system.

\end{figure}
The corresponding Hamiltonian is
\begin{equation}
    H_S = \sum_{i,j} \bm{S}_i \cdot\bm{D}^{ij}\cdot \bm{S}_j  + \sum_i \bm{B}\cdot \bm{\Gamma}^i \cdot \bm{S}_i
    \label{eq:5_spin_H}
\end{equation}
where 
\begin{equation*}
   \bm{D}^{ij} =  
   \begin{pmatrix}
        J_x^{ij} & K_z^{ij} & 0 \\
        -K_z^{ij} & J_y^{ij} & 0 \\
        0 & 0 & J_z^{ij}
    \end{pmatrix}    
\end{equation*}
is the tensor describing the interaction between the spins $i$ and $j$. For the antiferromagnetic interactions we set $J_x^{ij}=J_y^{ij}=J_z^{ij}$, with $J_x^{12}=J_x^{23}=J_x^{31}=0.3 \,\text{meV}$ and $J_x^{1m}=J_x^{2m}=J_x^{3m}=0.1 \,\text{meV}$ where $m=4,5$.
$K_z^{ij}=J_z^{ij}/10$ are the components along the $z-$axis of Dzyaloshinskii–Moriya interactions (DMI). For this system, we choose $B=1.0\,\text{T}$ and identical, diagonal, isotropic $\bm{\Gamma}^i$ tensors, with components $\Gamma_{xx}=\Gamma_{yy}=\Gamma_{zz}=2.2$ $\mu_B$, as in subsection \ref{subsec:singleGiant}.
To illustrate the results, among the 32 energy levels we selected the ones labeled as $\ket{1},\ket{3},\ket{9},\ket{14},\ket{21},\ket{26}$ because they allow us to create a wide set of superpositions with different values of $\Delta$ parameter.

\begin{figure}[h]
    \includegraphics[width=0.95\linewidth]{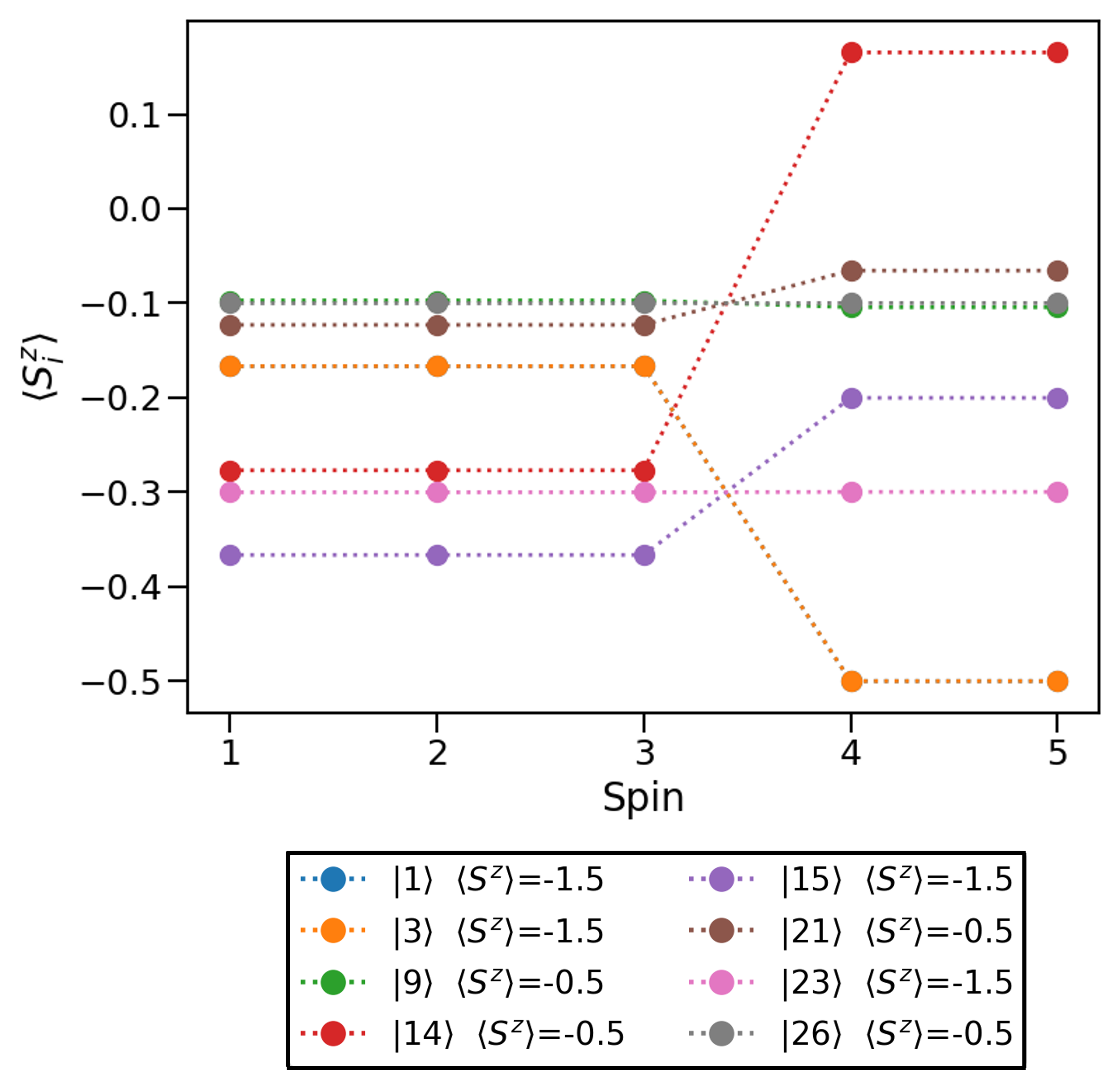}
    \caption{Expectation values of the $S_i^z$ operators for the eigenstates of the system considered in \ref{subsec:5_spins}. In the legend, we reported the label of the states and the value of $\braket{S^z}=\sum_i \braket{S_i^z}$.}
    \label{fig:5spins_texture}
\end{figure}

In figure \ref{fig:5spins_texture} we report the expectation values of the $S_i^z$ operators for the states considered above; only the $z$ component is shown because the other ones are almost zero. Note that $\braket{1|S_i^z|1}=\braket{3|S_i^z|3} \forall i=1,\dots,5$ while the curves associated to the states $\ket{9}$ and $\ket{26}$ are very similar but not identical.

\begin{figure}[h]
    \includegraphics[width=0.99\linewidth]{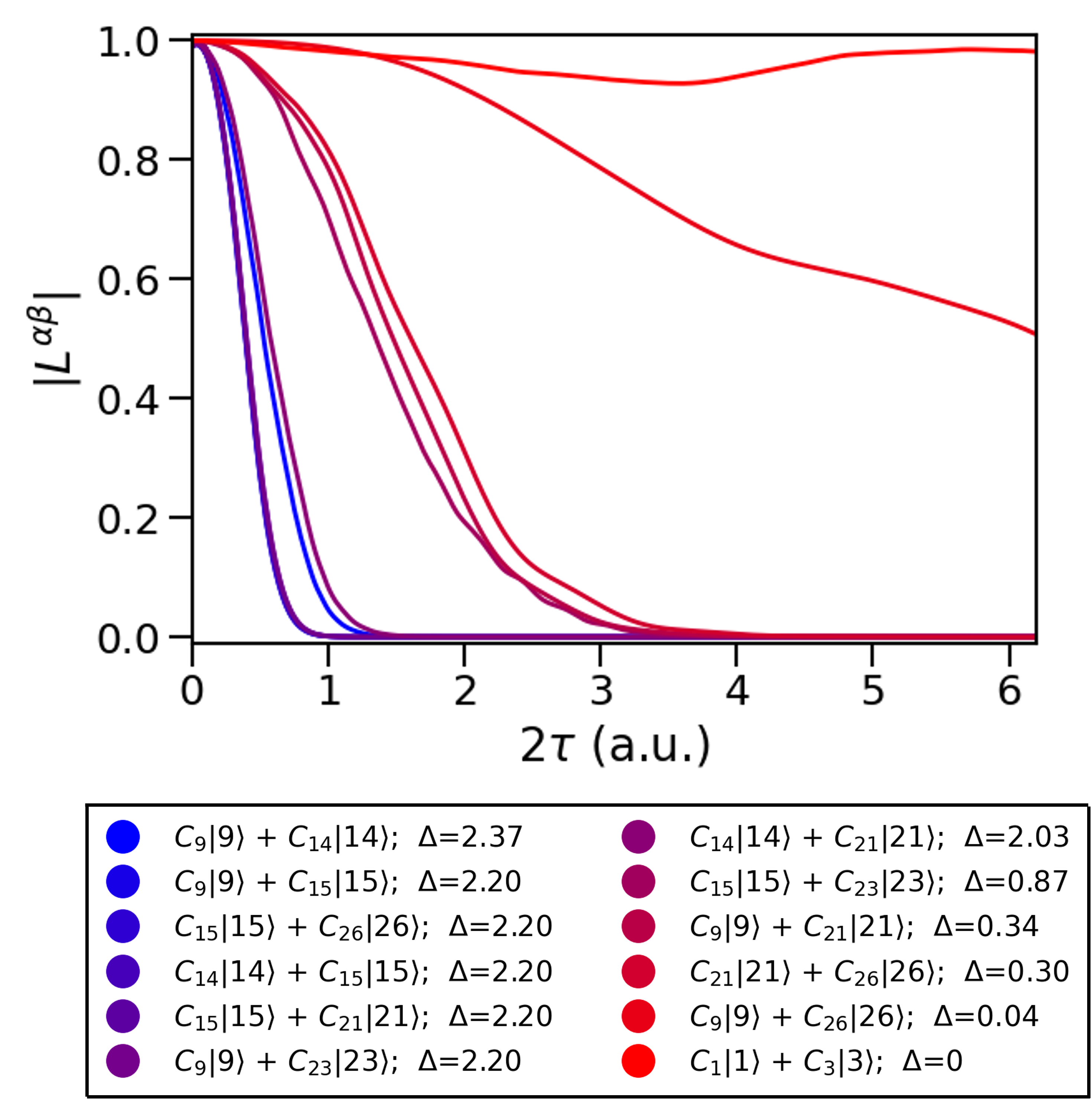}
    \caption{Coherence factor curves for the superposition of eigenstates illustrated in the legend. The value of the $\Delta$ parameter is also shown. The arbitrary unit is the time requested for the superposition $\ket{\psi}_{9,14}$ to reach the value of $0.001$ of the coherence factor.}
    \label{fig:5spins_coherence}
\end{figure}

In figure \ref{fig:5spins_coherence} we report the coherence factor as a function of time, the arbitrary unit is the time requested for the superposition $\ket{\psi}_{9,15}$ to reach the value of $0.001$ of the coherence factor (approximately $50\, \mu s$). For each state $\ket{\psi}_{\alpha,\beta}$  we report the corresponding value of $\Delta$ parameter. Note that, for the state $\ket{\psi}_{1,3}$, the curve fluctuates around the maximal value because $\Delta=0$, thus the term $\Tilde{R}$ in equation \ref{eq:expanded_comm} can not be neglected, as discussed in detail in supplementary material \ref{subsec:suppl_5_spin_R_term}.
Anyway, as outlined at the end of section \ref{subsec:neglect_sec_order}, this is the ideal condition that we would like to achieve.

Moreover, the results show that, for each superposition, the difference in \textit{total} spin $\braket{\beta |S^z|\beta}-\braket{\alpha |S^z|\alpha}$ does not explain the obtained coherence factor times.
Consider, for example, the eigenstates $\ket{9},\ket{21},\ket{26}$. They have the same expectation value for the operator $S^z$, as seen in the legend of figure \ref{fig:5spins_texture}. Still, $\ket{\psi}_{9,26}$ show completely different coherence factor time compared to the states $\ket{\psi}_{9,21}$ and $\ket{\psi}_{21,26}$, thus this set of eigenstates cannot be considered to design a 3-levels qudit.
These results can be completely explained only by comparing the expectation values of the local spin operators for the states $\ket{\alpha}$ and $\ket{\beta}$, involved in each superposition $\ket{\psi}_{\alpha,\beta}$. To this purpose, the considered $\Delta$ parameter estimates how far we are from the ideal condition where the coherence factor does not decrease.

Finally, note that the curves in figure \ref{fig:5spins_coherence} are not all ordered according the corresponding $\Delta$.
This result does not contrast with the theory developed above. Indeed, the $\Delta$ parameter defined in equation \ref{eq:spin_texture_simple} serves as a general measure associated with each pair of eigenstates and does not capture fine details.
For example, the curve corresponding to the state $\ket{\psi}_{9,14}$, with $\Delta=2.37$, decays slightly more slowly than the ones associated $\Delta=2.2$.
If the system contains more than one spin this measure is no longer a fine descriptor of the rate of decoherence, as happened for the one-spin system described in \ref{subsec:singleGiant}, and the curves do not strictly follow the order established by the parameter. Anyway, it statistically well describes the observed coherence factor decay, confirming that the smaller the value of $\Delta$, the longer the coherence time, in agreement with the theoretical results of section \ref{sec:theory}. 

\subsection{\label{subsec:qudit} Qudit proposal}
Equipped with the concepts learned through the examples in subsections \ref{subsec:singleGiant} and \ref{subsec:5_spins}, we started looking for a potential qudit candidate. Note that, to realize a \textit{qubit}, it is enough to find two eigenstates of a system that maximize the coherence factor time of the superposition. Instead, for a d-levels \textit{qudit}, there are $d(d-1)/2$ pairs of eigenstates, each one prepared in a superposition with a specific coherence factor time. It follows that a qudit can be designed only if all the coherence factor curves are similar in their temporal decay. As already demonstrated in the cases of the single giant spin and of the five spins system with competing interactions, long coherence factor times can be obtained by choosing eigenstates with the appropriate expectation value of the local spin operators. 

It must be emphasized that the search for a real molecular qudit candidate with the appropriate eigenstate structure is constrained not only by the requirement of minimal decoherence, but also by the practical limitations of implementation and manipulation.
For these reasons, we carefully selected eigenstates with the following features: the perturbative Schrieffer-Wolff transformation described in section \ref{subsec:ModelHamiltonian} must be valid, implying that the corresponding eigenvalues must not be nearly degenerate; these eigenvalues should fall within a range that can be experimentally addressed by electromagnetic pulses in the microwave spectrum to ensure full connectivity between states; the magnetic matrix elements for transitions between them should be sufficiently large to enable qudit manipulation within a reasonable timescale; and finally, the selected states should not be too far from the ground state to minimize relaxation due to phonon emission.

To increase the number of states satisfying the constraints described above, we propose a system composed of six $S=1/2$ spins (see figure \ref{fig:frustratedQudit}), obtained by adding a spin to the center of the triangular basis of the structure illustrated in figure \ref{fig:frustrated_system}. The corresponding Hamiltonian is shown in equation \ref{eq:5_spin_H}. 
Optimal qudits are obtained when the Zeeman term and the AFM exchange couplings are set to comparable values ranging from $0.1\,\text{meV}$ to $1\,\text{meV}$, with the DMI approximately one tenth of these values. For instance, a possible choice is to apply a magnetic field slightly tilted away from the 
$z$-axis to enhance the connectivity between eigenstates. Specifically, we set $B=1.0\,\text{T}$, tilted by $\pi/18\,\text{rad}$ around the y-axis. Moreover, we supposed values of the exchange parameters to vary approximately from $0.5\,\text{meV}$ to $0.1\,\text{meV}$, as detailed in supplementary information \ref{subsec:suppl_param_qudit}.
Highly symmetric molecular structures are usually imperfect and often exhibit distortions. To avoid results that depend strictly on perfect symmetry, we introduce slight variations in the parameters to break it. 
The same random variation is applied to the gyromagnetic ratio tensors.

To single out the contributions of the eigenstate structure to decoherence, all the simulations have been performed considering the same bath used for the system described in section \ref{subsec:5_spins}. 
In order to highlight the effect of interactions, we have considered an isostructural system with uncoupled spins, thus setting the $D^{ij}$ tensor in equation \ref{eq:Hs} equal to zero $\forall\,i,j$. The corresponding temporal decay is shown as a black curve in figure \ref{fig:qudit_antif_coherence}. The time instant at which the coherence factor assumes the value of 0.001 is taken as a reference to set the timescale, which is approximately $30\, \mu s$.

\begin{figure}[]
    \includegraphics[width=0.99\linewidth]{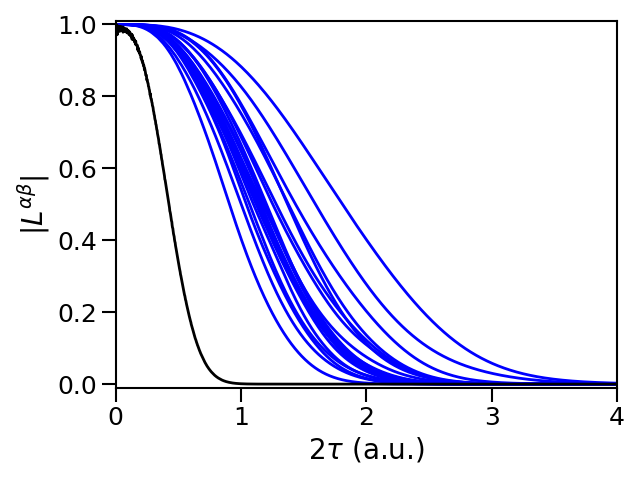}
    \caption{The 21 coherence factor curves for the 7 eigenstates with $S^z\approx0$ of the AFM system (in blue). The uncoupled-spins curve, in black, is used to define the time scale arbitrary unit.}
    \label{fig:qudit_antif_coherence}
\end{figure}

We selected 7 eigenstates with $S^z\approx0$, more details on the expectation values of the spin operators are given in supplementary material \ref{subsec:suppl_exp_values_spin_op}. We calculated the coherence factor for each of the corresponding 21 pair superpositions and reported the results with blue curves in figure \ref{fig:qudit_antif_coherence}. Note that the coherence factor curves for all 21 superpositions show longer coherence factor times with respect to the reference non-interacting system.
Moreover, the molecule exhibits full connectivity among all the levels, as the transition dipole moment for each pair of states is sufficiently large to enable transitions within a limited time.
This nontrivial feature is highly advantageous for fully harnessing the potential of qudits, and is essential for implementing certain quantum algorithms. For instance, some quantum error correction algorithms designed for qudits, require an all-to-all connection among levels.
We stress that this last system does not represent the optimal example of molecular spin qudit. In principle, coherence times could be significantly extended through the optimization of the Hamiltonian parameters and the appropriate selection of the states. 
However, this system can be seen as a prototypical example that embodies all the ingredients for the theoretical design of molecular spin qudits.

\section{\label{sec:conclusions} Conclusions}
In this work, we addressed the problem of decoherence in molecular spin systems at low temperatures. In particular, we focused on the pure dephasing resulting from the interaction of the central system with the surrounding nuclear spins bath. To do so, we developed a microscopic model of decoherence and implemented the CCE method to simulate the coherence decay over time numerically.  
This approach allowed us to demonstrate a key result: the coherence of a superposition of two eigenstates $\ket{\alpha}$ and $\ket{\beta}$ is preserved if and only if the two Hamiltonian operators that describe the associated conditional bath dynamics commute with each other. For molecular spin systems, this condition translates into requiring identical expectation values of local spin operators for the two eigenstates.

While this ideal condition is rarely met in real systems, we introduced a parameter $\Delta$ to quantify the deviations and we explored its impact on coherence times. Through numerical simulations, we validated our theoretical framework and demonstrated its applicability to systems ranging from single giant spins to composite structures with competing interactions.
These insights allowed us to define a recipe for the theoretical design of molecular spin qudits.  As a paradigmatic example, we proposed a composite system made of six spins 1/2, interacting with each other through antiferromagnetic exchange interactions. 
We showed that it is possible to exploit competing interactions to create a viable qudit candidate, with many low-energy states characterized by full connectivity. This is crucial for implementing quantum error correction schemes and other quantum algorithms, making such systems highly advantageous for quantum technologies. While the proposed system represents a significant step forward, it is important to note that it does not yet constitute the optimal molecular spin qudit. In principle, further optimization of Hamiltonian parameters and state selection could extend coherence times and improve qudit performance. However, this system serves as a prototypical example to demonstrate how chemical tunability of spin interactions can mitigate decoherence, offering an interesting perspective for more robust quantum information processing platforms.
The strategy developed here can be exploited in synergy with the mature approaches based on engineering the bath to push beyond the current limitations of spin qudits.\\
In conclusion, this work advances our understanding of decoherence in molecular spin systems and provides a practical framework for designing and optimizing molecular spin qudits, bringing them closer to practical implementations in quantum information processing.\\

{\bf Acknowledgments}
This work received financial support from the European Union-NextGenerationEU, PNRR MUR projectPE0000023-NQSTI.

\bibliography{bibliography}

\newpage
\onecolumngrid
\section*{Supplementary Information}
\subsection{Neglecting the second order perturbed terms}
\label{subsec:suppl_neglect}
Here, we will show how and why we can neglect all the commutators involving the $H_{SB2}$ terms in equation \ref{eq:full_commutator}. Without loss of generality, we suppose a magnetic field acting along the z-axis.
Consider the last two terms in the Hamiltonian of equation \ref{eq:SWH}. We can estimate these interactions, due to a bath spin located position $\mathbf{r}$, as follows
\begin{equation}
    \begin{split}
        & H_{SB1}(\mathbf{r})\approx m^zA(\mathbf{r})\\
        & H_{SB2}(\mathbf{r})\approx 2\frac{(m^z)^2}{\Delta E}A(\mathbf{r})  \sum_{\mathbf{d}} A(\mathbf{d}) \\
    \end{split}
\end{equation}
where $A$ is the mean interaction due to spin in position $\mathbf{d}$, $\Delta E$ is a typical energy difference between two system eigenstates and $m^z$ is the eigenvalue of the operator $S^z$ relative to $\ket{\psi}$, i.e. the eigenstate we are taking into account. Considering equation \ref{eq:A}, we infer that
\begin{equation}
    A(\mathbf{r})=\frac{\mu_0\hbar^2}{4\pi\abs{\mathbf{r}}^3}\Gamma\gamma
\end{equation}
and, by assuming a continuous and uniform distribution of spins into a sphere centered around the system, we find the following ratio 
\begin{equation}
   \frac{H_{SB2}(\mathbf{r})}{H_{SB1}(\mathbf{r})}\approx\frac{m^z\mu_0\hbar^2\Gamma\gamma}{2\Delta E}\big(\frac{1}{R_{min}^2}-\frac{1}{R_{max}^2} \big)
\end{equation}
where $R_{min}$ and $R_{max}$ are respectively the minimal and the maximal radius for the sphere. The maximum value of this function ($R_{max}=\infty$) is 
\begin{equation}
    \frac{H_{SB2}(\mathbf{r})}{H_{SB1}(\mathbf{r})}\approx10^{-2} 
\end{equation}
where we assumed $R_{min}=3\mathring{A}$ and typical values for $\Delta E$, $m^z$, $\gamma$ and $\Gamma$. Thus, $H_{SB2}(\mathbf{r})$ is negligible with respect to $H_{SB1}(\mathbf{r})$ for each bath spin, regardless their position.
It follows that the last commutator in equation \ref{eq:full_commutator} can be neglected with respect to the fourth and the fifth, which can be neglected with respect to the first one. Finally, the third commutator is smaller than the second one, leading to equation \ref{eq:approx_commutator}.

If the terms $(\bra{\beta}S_k^{\mu}\ket{\beta}-\bra{\alpha}S_k^{\mu}\ket{\alpha})\Gamma_{\eta\mu}^k$ in equation \ref{eq:expanded_comm} are small and comparable to $\Tilde{R}$, equation \ref{eq:approx_commutator} is no longer valid and we can approximate equation \ref{eq:full_commutator} as follows.
For each pair of spins ($i,j$) belonging to the bath with components ($\mu,\nu$), we can compare the strength of the interactions due to $\mathbf{J}$ and $\mathbf{T}$ terms (see equations \ref{eq:J} and \ref{eq:T}, respectively). We define the tensor $\mathbf{\Lambda}$ as
\begin{equation}
    \Lambda^{ij}_{\mu \nu}=\abs{\frac{T_{\mu \nu}^{ij}}{J_{\mu \nu}^{ij}}}
\end{equation}
Furthermore, we associate the following weight $\mathbf{P}$ to each coordinate 
\begin{equation}
P^{\mu \nu}_{ij}=\frac{\abs{J_{\mu \nu}^{ij}}}{\sum_{i,j,\mu,\nu}\abs{J_{\mu \nu}^{ij}}}
\end{equation}
that is normalized by definition and becomes greater and greater as the interaction $J_{\mu \nu}^{ij}$ increases.
The mean value of $\mathbf{\Lambda}$ can be evaluated as
\begin{equation}
    \braket{\mathbf{\Lambda}}=\sum_{i,j,\mu,\nu}\Lambda^{ij}_{\mu \nu}P^{\mu \nu}_{ij}=\frac{\sum_{i,j,\mu,\nu}\abs{T_{\mu \nu}^{ij}}}{\sum_{i,j,\mu,\nu}\abs{J_{\mu \nu}^{ij}}}
    \label{eq:lambda_gen}
\end{equation}
that is equivalent to calculating the ratio between the sum of the absolute value of the interactions. We can estimate this mean value for typical systems by assuming a continuous distribution of spins into a sphere of radius $R_{max}$ and obtain
\begin{equation}
    \braket{\mathbf{\Lambda}}=\frac{\mu_0(\Gamma\hbar m^z)^2}{8\Delta E}\frac{(R_{max}^2-R_{min}^2)}{R_{max}^4R_{min}^4}\frac{1}{I}
\end{equation}
with 
\begin{equation}
\begin{split}
    & I=  \int_{R_{min}}^{R_{max}}\int_{R_{min}}^{R_{max}}\int_{t(r1_,r2,l)}^{\pi} \frac{dr_1 dr_2 d\theta}{\big(r_1^2 +r_2^2 -2r_1r_2\cos\theta \big)^{3/2}} \\
    & t(r_1,r_2,l)=\arccos\big(\frac{r_1^2+r_2^2-l^2}{2r_1r_2}\big)
\end{split} 
\end{equation}
where $l$ is the minimum distance between the bath's spins. 
By keeping $l$ and $R_{min}$ fixed, we see that the value of $\braket{\mathbf{\Lambda}}$ decreases while $R_{max}$ increases as expected by considering the interactions involved in equation \ref{eq:lambda_gen}.
This consideration appears evident in figure \ref{fig:lambda}, where we assumed $R_{min}=3\mathring{A}$, $l=3\mathring{A}$ and typical values for the other parameters.
\begin{figure}[h]
    \includegraphics[width=0.5\linewidth]{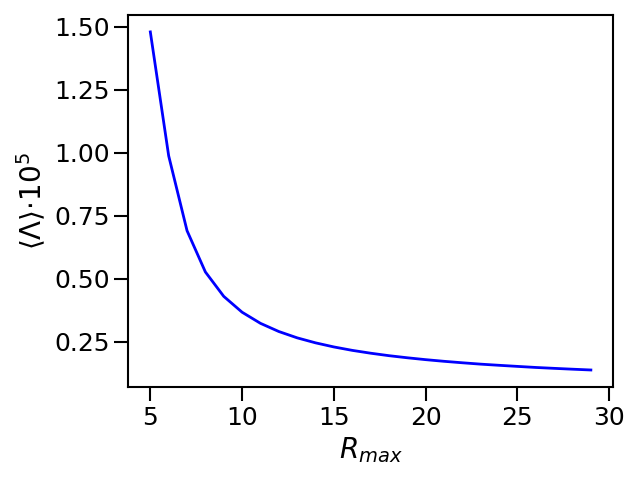}
    \caption{Expectation value of $\Lambda$ when the parameters assume the typical values encountered in MNMs.}
    \label{fig:lambda}
\end{figure}
Moreover, it can be noticed that the values of $\braket{\mathbf{\Lambda}}$ are $O(10^{-5})$, allowing us to neglect the fourth and fifth term with respect to the second one in equation \ref{eq:full_commutator}, as well as the sixth one with respect to the third. In this case, equation \ref{eq:full_commutator} reduces to equation \ref{eq:approx_comm_2}, that we report here 
\begin{equation}
    [H^{\alpha},H^{\beta}] \approx \sum_{\substack{k,\mu,\eta}} (\bra{\beta}S_k^{\mu}\ket{\beta}-\bra{\alpha}S_k^{\mu}\ket{\alpha})\Gamma_{\eta\mu}^k O_{\eta}^k+[H_B,\Delta H_{SB2}]
\end{equation}
As outlined in the main text, nevertheless the differences $(\bra{\beta}S_k^{\mu}\ket{\beta}-\bra{\alpha}S_k^{\mu}\ket{\alpha})\Gamma_{\eta\mu}^k$ are no more the crucial quantity to relate the coherence factor to the eigenstates of the system, we achieved the ideal condition corresponding to long coherence times because the term $\Delta H_{SB2}$ is small by definition and, as a consequence, $[H^{\alpha},H^{\beta}] \approx 0$.

\subsection{Numerical results and second order perturbed terms}
\label{subsec:suppl_5_spin_R_term}
Here we investigate the contribution of the $\Tilde{R}$ term to the coherence factor curves obtained for the system described in section \ref{subsec:5_spins}. In figure \ref{fig:R_term} we report in blue the curves shown in figure \ref{fig:5spins_coherence} of the main text and, in red, the ones obtained by removing the terms defined in equation \ref{eq:T} from the Hamiltonian \ref{eq:SWH}.
In the latter case, the term $\Tilde{R}$ in equation \ref{eq:expanded_comm} is zero and the relation \ref{eq:approx_commutator} is exact. 
We note that, for each pair of eigenstates, the faster they decay, the smaller the difference between the red and the blue curves: they are indistinguishable for each superposition except for the two with longer coherence times.
By considering the discussion made in section \ref{subsec:neglect_sec_order}, this is exactly what we expect: the more the differences between the expectation values of the spin operator become small ($\Delta$ parameter decrease), the more the coherence times become longer and the second order terms non-negligible because comparable with the leading terms.

\begin{figure}[h]
    \includegraphics[width=0.6\linewidth]{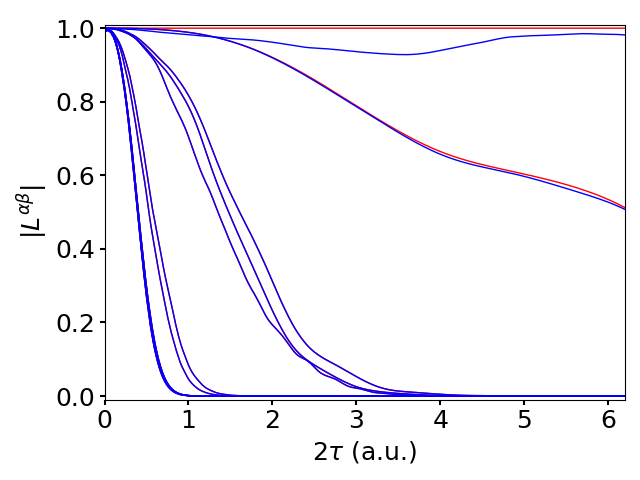}
    \caption{For the system described in section \ref{subsec:5_spins}, we report the curves associated with a set of eigenstates: in blue, the one showed in figure \ref{fig:5spins_coherence}, in red these eigenstates are computed by removing the terms \ref{eq:T} from the Hamiltonian.}
    \label{fig:R_term}
\end{figure}

\subsection{Parameters for the Hamiltonian of qudit proposal}
\label{subsec:suppl_param_qudit}
Here we report the values of all the parameters for the system described in \ref{subsec:qudit}.
For the interaction between the spins we have:

$J^{ij}_x=J^{ij}_y=J^{ij}_z$, $J^{12}_x=0.5 \,meV$, $J^{23}_x=1.01 J^{12}_x$, $J^{31}_x=1.08 J^{12}_x$, $J^{14}_x=0.1 \,meV$,
$J^{24}_x=0.95 J^{14}_x$, $J^{34}_x=1.03 J^{14}_x$, $J^{15}_x=1.10 J^{14}_x$, $J^{25}_x=0.89 J^{14}_x$, $J^{35}_x=0.98 J^{14}_x$, $J^{45}_x=0.1 J^{12}_x$, $J^{16}_x=1.1 J^{12}_x$, $J^{26}_x=1.05 J^{16}_x$, $J^{36}_x=0.93 J^{16}_x$, $J^{46}_x=1.05 J^{14}_x$, $J^{56}_x=0.92 J^{46}_x$, $K^{ij}_z=J^{ij}_z/10$

As stated in the main text, $B=1 \,T$ and tilted by $\pi/18$ rad around the y-axis. We consider the following diagonal and isotropic gyromagnetic ratio tensors:
$\Gamma_{xx}^1=2.210 \,\mu_B$, $\Gamma_{xx}^2=2.200 \,\mu_B$, $\Gamma_{xx}^3=2.180 \,\mu_B$, $\Gamma_{xx}^4=2.190 \,\mu_B$, $\Gamma_{xx}^5=2.205 \,\mu_B$, $\Gamma_{xx}^6=2.195\,\mu_B$

\subsection{Expectation values of spin operators for qudit proposal}
\label{subsec:suppl_exp_values_spin_op}
In the following figures (\ref{fig:Sx},\ref{fig:Sy},\ref{fig:Sz}), we report the expectation values of the local spin operators for the qudit proposal defined in section \ref{subsec:qudit}. In the legend of each figure, we show the label of the chosen eigenstates.
\begin{figure}[h]
    \includegraphics[width=0.8\linewidth]{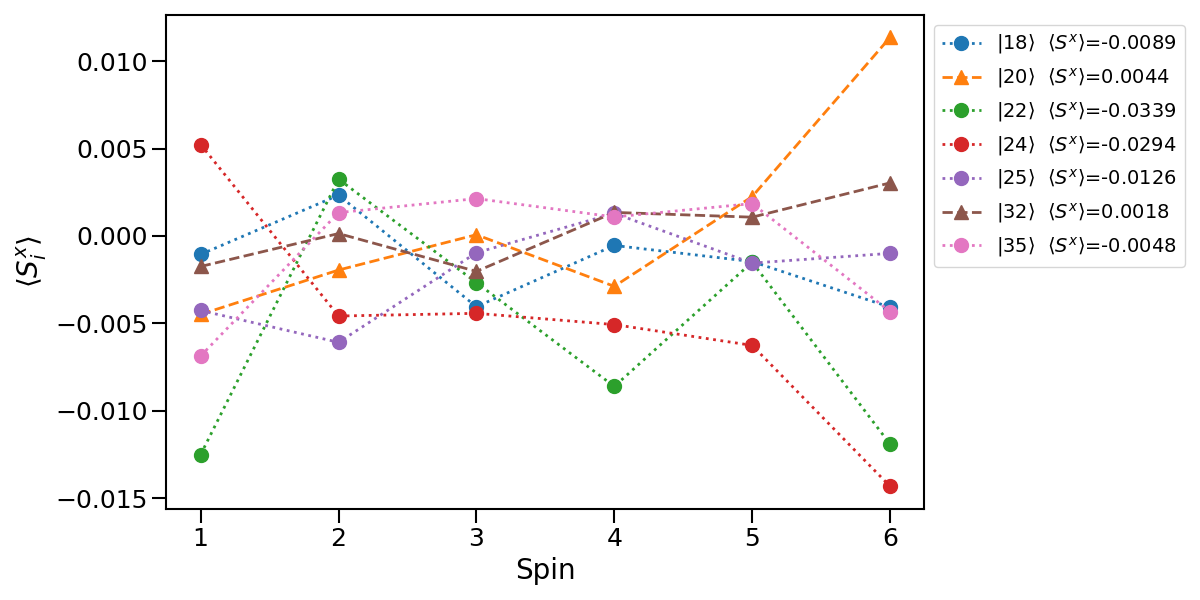}
    \caption{Expectation values of the $S^x_i$ operators. The values of $\braket{S^x}=\sum_i \braket{S_i^x}$ are reported in the legend.}
    \label{fig:Sx}
\end{figure}
\begin{figure}[h]
    \includegraphics[width=0.8\linewidth]{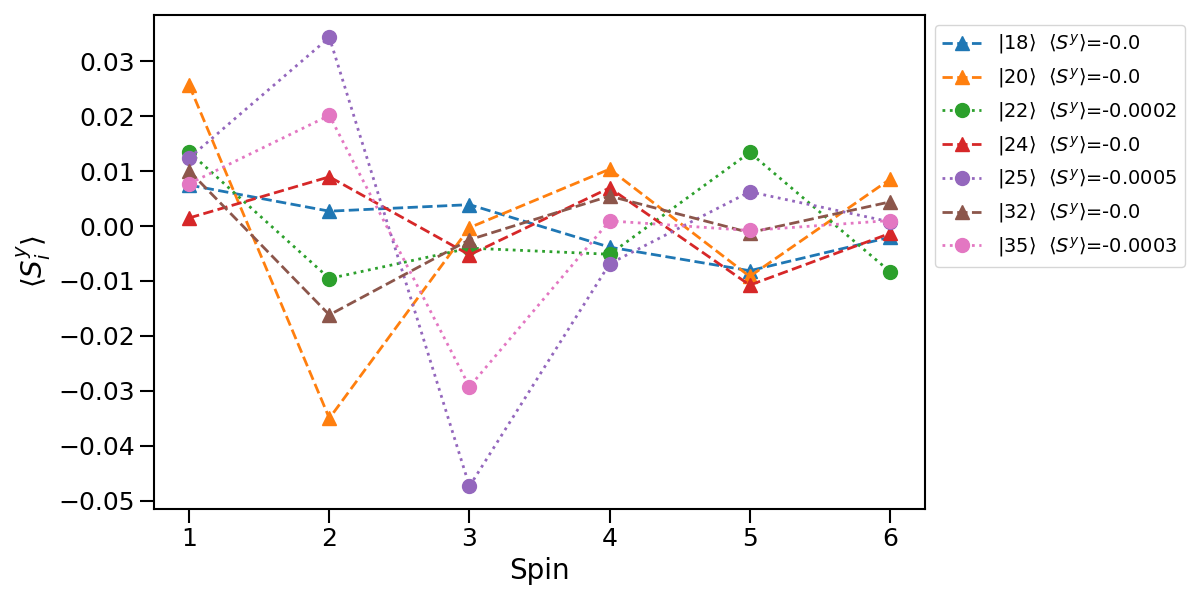}
    \caption{Expectation values of the $S^y_i$ operators. The values of $\braket{S^y}=\sum_i \braket{S_i^y}$ are reported in the legend.}
    \label{fig:Sy}
\end{figure}
\begin{figure}[h]
    \includegraphics[width=0.8\linewidth]{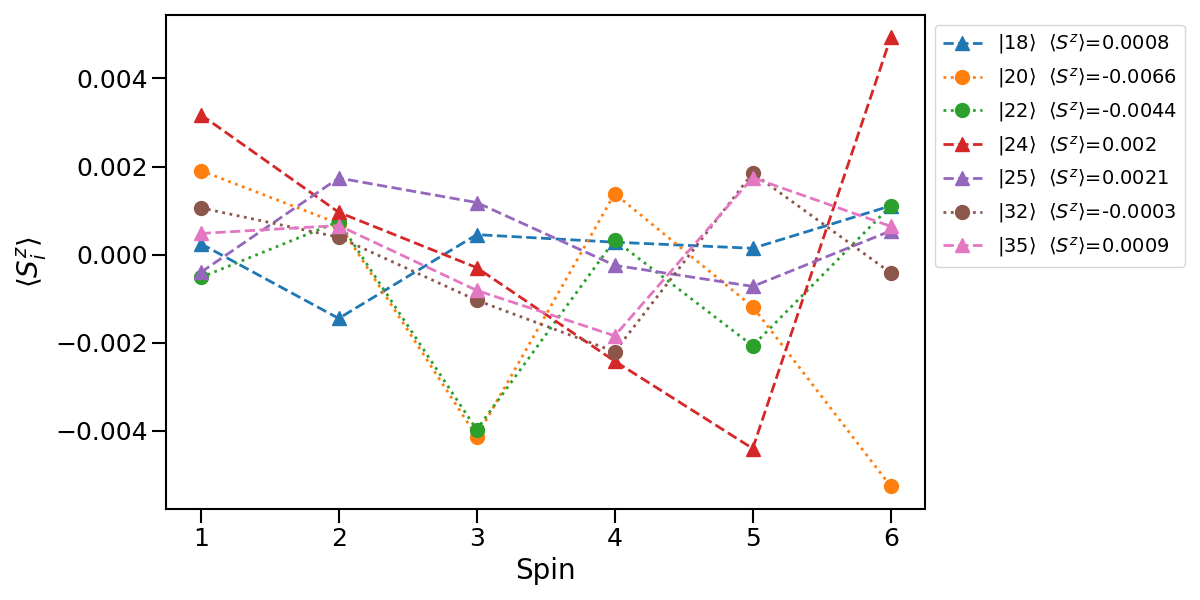}
    \caption{Expectation values of the $S^z_i$ operators. The values of $\braket{S^z}=\sum_i \braket{S_i^z}$ are reported in the legend.}
    \label{fig:Sz}
\end{figure}
\end{document}